**Title: Molecular structure elucidation with charge-state control**

**Authors:** Shadi Fatayer[1]*, Florian Albrecht[1], Yunlong Zhang[2], Darius Urbonas[1], Diego Peña[3], Nikolaj Moll[1], Leo Gross[1]*

**Affiliations:**

[1]IBM Research – Zurich.

[2]ExxonMobil Research and Engineering Company.

[3]Centro Singular de Investigación en Química Biolóxica e Materiais Moleculares (CIQUS) and Departamento de Química Orgánica, Universidade de Santiago de Compostela, 15782 Santiago de Compostela, Spain.

*Correspondence to: sfa@zurich.ibm.com, lgr@zurich.ibm.com

**Abstract:**

The charge state of a molecule governs its physicochemical properties, such as conformation, reactivity and aromaticity, with implications for on-surface synthesis, catalysis, photo conversion and applications in molecular electronics. On insulating, multilayer NaCl films we control the charge state of organic molecules and resolve their structures in neutral, cationic, anionic and dianionic states by atomic force microscopy, obtaining atomic resolution and bond-order discrimination using CO functionalized tips. We detect changes in conformation, adsorption geometry and bond-order relations for azobenzene, tetracyanoquinodimethane and pentacene in multiple charge states. Moreover, for porphine we investigate the charge-state-dependent change of aromaticity and conjugation pathway in the macrocycle. This work opens the way to studying chemical-structural changes of individual molecules for a wide range of charge states.

**One Sentence Summary:** Adsorption properties and bond-order relations of single molecules in multiple oxidation states are resolved by atomic force microscopy.

**Main Text:**

Structure and function of a single molecule can be drastically altered based on its oxidation state (*1*). The charge-induced changes of a molecule have fundamental implications for chemical reactions, catalysis, electrochemistry, photoconversion and charge transport. Charged molecules can be studied in various environments, e.g. gas-phase, solution and solid. Experimentally, structural information is mostly obtained from X-ray diffraction for molecular solids (*2*), vibrational and optical spectroscopy for molecules in solution and gas-phase (*3*). Typically, counterions are used to stabilize charged molecules and avoid fragmentation or charge leakage in solids and solutions (*4*). Hence, the crystallization and the counterions may affect the geometry of the molecules. The aforementioned experimental techniques predominantly probe a large number of molecules. The ability to detect single-molecule changes upon charging with atomic resolution would allow new phenomena to be observed (*5*) and the influence of the environment to be quantified. Moreover, it would also impact applications such as single-electron-transfer based devices as well as thin film devices and provide fundamental insights into redox-reactions and charge carrier injection.

Scanning tunneling microscopy (STM) and atomic force microscopy (AFM) imaging of adsorbed molecules in different charge states has been shown in specific cases on ultrathin insulating films, e.g., bilayer NaCl, where tunneling to the substrate and STM operation is possible. For some molecules, two charge states can be stabilized under special conditions of energy level alignment and relaxation energy, i.e., when reorganization in the film/molecule leads to level-shifting across the Fermi level. In such special cases, molecular conformational switching (*6*), tunnel barrier changes (*7*) and formation of a metal-molecule complex (*8*) have been demonstrated. On thicker insulating films electrons cannot be exchanged with the substrate and thus multiple charge states can be stabilized in general (*9*) and reorganization energies determined (*10*). Here, we complement single-electron sensitivity (*11–13*) and charge-state control (*9*) with atomic resolution of AFM by using CO tips (*14*). We resolve molecules in multiple charge states (also called oxidation states) with atomic resolution including bond-order analysis. The molecular charge states are controlled by deliberately exchanging electrons between tip and molecule on an insulating substrate (*9*), accessing multiple charge states including doubly charged ions. Surprisingly, the changes in the molecular geometry of charged species can often be rationalized employing basic analysis in organic chemistry such as the evaluation of Lewis resonance structures or Clar's sextet rule.

Here, we investigate four model compounds with different properties and applications related to their charge-state transitions. We resolve minute changes in adsorption geometry, bond-order relations and aromaticity. All the experiments were performed on a NaCl film thicker than 20 monolayer (ML) on Cu(111) using a qPlus-based (*15*) AFM under ultra-high vacuum at 5 K (see Methods and Materials for details). The CO-functionalized tip was approached with frequency-shift ($\Delta f$) feedback to the multilayer NaCl area, where constant $\Delta f$ AFM imaging was used to locate molecules. To identify and manipulate the charge state of molecules we used Kelvin probe force spectroscopy, i.e. $\Delta f$ as a function of sample voltage $V$. Transitions in the molecular charge state were observed as steps between different $\Delta f(V)$ parabolas (*11–13*), where each step upon ramping $V$ in the positive (negative) direction corresponds to a decrease (increase) in charge state by attaching one electron (hole). The charge states were assigned by comparison to the ion resonances and pristine charge states on NaCl(2ML)/Cu(111) employing the scattering of surface-state electrons and/or interface-state localization as indications (*16*). Moreover, the separation (gap) between transition voltages on the insulating film was another indicator (Fig. S8). To avoid exchange of electrons with the tip during imaging, we obtained AFM images at voltages between different charge transitions, i.e., in different charge states. The AFM images were taken in constant-height mode, at a tip height $z$ near the minimum of $\Delta f(z)$ above the molecule, which roughly corresponds to a distance of 3.9 Å between the molecule imaged and the oxygen atom of the tip (*17*). The atomic contrast of the AFM images confirms that the tip remained CO functionalized, even though several volts are applied.

Azobenzene (*A*), see Fig. 1A, is an archetypical molecular electronics building-block (*18*) and photomechanical actuator (*19*). *A* can be switched between *cis* and *trans* isomer by light (*20*, *21*), tunneling electrons (*22*) and electric field (*23*). The AFM image of neutral $A^0$, displayed in Fig. 1B, shows that it is adsorbed with both phenyl rings slightly tilted out of plane with respect to the NaCl surface. The phenyl rings are approximately parallel (Fig. S2-S4), indicated by the same sides of the phenyl rings appearing bright and with comparable $\Delta f$ contrast in the AFM image. An electron was attached to $A^0$ when $V$ was ramped above 2 V (Fig. 1A), creating the anion $A^{-1}$. In $A^{-1}$ (Fig. 1C) the phenyl rings are also tilted out-of-plane, but in opposite directions, indicated by opposite sides of the two phenyl rings appearing bright in the AFM image (Fig. S5). Hence, by reducing $A^0$ the conformation of the molecule changes. A sequence of AFM images of $A$ while alternating its charge state demonstrates the reversibility and reproducibility of this charge/conformation switching (Fig. S6). The AFM measurements are in excellent agreement with AFM simulations (*24*) of the adsorbed molecule (Fig. 1D and 1E) performed in the respective geometries obtained by density functional theory (DFT) calculations (see supplemental material for details and Table S1 and S2) of $A^0$ (Fig. 1F) and $A^{-1}$ (Fig. 1H). We find that for both oxidation states $A$ is in the *trans* conformation, however, there are small differences in the molecular geometry. $A^0$ is planar and the molecular plane is tilted by 17 degrees with respect to the surface plane. Conversely, $A^{-1}$ is non-planar with the phenyl rings tilted in opposite directions by about 4 degrees with respect to the surface plane. The switch from a planar to a non-planar conformation can be rationalized by the reduction of the azo group (N=N) when switching from $A^0$ to $A^{-1}$, which alters the π conjugated system and induces the distortion from planarity. The example of azobenzene demonstrates that charge-induced changes in the adsorption geometry and molecular conformation, i.e., tilts of a few degrees of molecular moieties, can be detected.

Next, we discuss the possibility of resolving charge-induced changes in bond-order relations. Bond-order analysis is a powerful tool for resolving structural and electronic properties of molecules. In general, bond-order differences can be qualitatively resolved by AFM as different brightness (Δ$f$ values) and as different apparent bond length (*17*, *25*). Brighter Δ$f$ contrast is caused by increased Pauli repulsion due to increased electron density and thus indicates smaller bond length. At small tip-sample distances the tilting of the CO at the tip strongly affects the images and leads to a magnifying effect, increasing bond-length differences by about one order of magnitude (*17*). Moreover, background forces related to the local potential landscape contribute to the CO-tilting and thus also affect the apparent bond length (*17*, *24*). Therefore, one should only compare bonds with similar local environments. Before applying this method to resolve bond-order relations of molecules with oxidation-state-dependent functions, we first characterize the well-studied model system of pentacene ($P$).

The Δ$f(V)$ spectrum of $P$ (Fig. S8) reveals four different oxidation states, cation ($P^{+1}$), neutral ($P^0$), anion ($P^{-1}$) and dianion ($P^{-2}$). Differences can be observed in their AFM images, as shown in Fig. 2A-D and in Fig. S9. We observe an apparent contraction (elongation) of the molecule along its short (long) axis with increased negative charge which agrees with the calculations (Fig. S10). However, one must be cautious when comparing apparent bond lengths obtained at different voltages. The electric field changes with different $V$ and this will have an impact on the spring constant and tilt of the CO molecule at the tip apex (*24*). In fact, we observe an overall compression of the molecular image of $P^0$ with increased $V$ (Fig. S11 and SM1). However, differences between individual bonds within one image, i.e., measured at identical voltages, can be compared without such systematic error. In the following, we compare only the contrast within individual AFM images and bonds with a similar local environment.

Within the images of $P^{-1}$ and $P^{-2}$ we observe a Δ$f$ modulation, with the centers of the second and fourth ring appearing darker than the centers of the other rings, indicating that these rings are of increased diameter and/or exhibit a reduced electron density. DFT calculations of the average C-C bond length within individual rings (Fig. 2E) show that $P^{-1}$ and $P^{-2}$ feature second and fourth rings with an increased averaged bond length compared to the other rings. For the dianion $P^{-2}$ the effect can be rationalized by considering the reduction of pentacene to form radical anions in the second and fourth rings (Fig. 2F), leading to a structure with three Clar aromatic sextets distributed in the first, third and fifth ring. The measurements of pentacene showcase that charge-induced small effects in the bond-length relations are resolved by AFM. Importantly for redox reactions, we obtain the locations of sites with increased radical anion character upon charging.

Having corroborated charge-state control and detection of charge-induced changes in adsorption geometry and bond-order relations, we apply our method to the electron acceptor tetracyanoquinodimethane ($T$) (Fig. 3A), relevant for doping in organic electronics (*26*). The $\Delta f(V)$ spectrum of $T$ (Fig. S14) reveals three different oxidation states, neutral ($T^0$), anion ($T^{-1}$) and dianion ($T^{-2}$). Consistent with its large electron affinity we observe the anion $T^{-1}$ at 0 V already. The AFM image of $T^0$ (Fig. 3B) shows four attractive lobes, two large ones and two small ones (Fig. S15), which we assign to an upstanding adsorption conformation (*27*) for $T^0$. AFM measurements of $T^0$ at smaller tip-molecule distances than in Fig. 3B resulted in the molecule being picked up by the tip. In contrast, the AFM image of $T^{-1}$ (Fig. 3C) resolves the central carbon ring adsorbed parallel to the surface. The adsorption orientation switch between $T^0$ and $T^{-1}$ is reversible and does not result in lateral movement of the molecule (Fig. S16). The contrast on the central ring of $T^{-1}$ indicates bond length alternation (BLA). Increased $\Delta f$, indicative of shorter bond length, is observed above the double bonds of the ring as depicted in its neutral chemical structure (Fig. 3A). The BLA in the ring suggests that $T^{-1}$ does not possess a perfect benzenoid character (*28*). However, in $T^{-2}$ the $\Delta f$ contrast of the central ring becomes homogeneous (Fig. 3D), indicating no BLA and thus a change to a benzenoid character (Fig. 3E), in agreement to previous calculations (*29*). In addition, comparing $T^{-2}$ to $T^{-1}$, the regions of the carbons connecting the cyano groups show increased $\Delta f$ for the dianion. Possible explanations are higher electron density at these carbons or adsorption-height changes in the termination of the molecule upon charge-state change. DFT calculations (Table S7 and S8) indicate that the cyano groups are more bent for $T^{-1}$ than for $T^{-2}$. In summary, for $T$ we observed adsorption orientation change upon reduction of $T^0$. Moreover, the prevalent phenyl character changes from quinoid in $T^{-1}$ to benzenoid in $T^{-2}$. This change is accompanied by small geometry changes in the dicyano moieties.

Lastly, we apply our method to porphyrins, which in multiple oxidation states fulfill essential functions in medicine, biology, chemistry and physics (*30*). Although controversial (*31*), the aromaticity and conjugation pathway of porphyrins shall also be oxidation-state-dependent. The parent compound of porphyrins is porphine ($F$), a fully conjugated substituent-free planar macrocycle (Fig. 4A). According to the [18]annulene model, the neutral molecule has an aromatic conjugation pathway involving 18π electrons (4n+2), indicated by the red colored bonds of the resonance structures shown in Fig. 4A, bypassing the NH in the pyrroles and the outer CH=CH groups of the azafulvene rings. Interestingly, upon double reduction the macrocyclic conjugation pathway changes to a formally antiaromatic 20π-electron (4n) system, encompassing the whole periphery of porphine (Fig. 4B). The change in the macrocyclic π conjugation can influence the global aromaticity of the molecule, although local heterocyclic π circuits (pyrrole subunits, 6π–electron, 4n+2) would contribute as well (*31*).

The $\Delta f(V)$ spectrum of $F$ (Fig. S20) reveals three different oxidation states, neutral ($F^0$), anionic ($F^{-1}$) and dianionic ($F^{-2}$). The AFM images and the respective Laplace-filtered images are shown in Fig. 4C-H and in Fig. S21 and S22. The location of the hydrogens inside the cavity can be inferred from AFM images at increased tip-sample distance (Fig. S23) and is indicated in Fig. 4C. We observed tautomerization switching (*32*) of $\boldsymbol{F}$ only at voltages larger than 3.7 V (Fig. S27). In the individual AFM images we observed apparent bond length differences in the pyrrole and azafulvene rings as well as in the methine bridges, which connect the 5-membered rings. The outer C-C bonds of the pyrrole and the azafulvene ring, labelled *a* and *c* (Fig. 4I), respectively, can be compared within an image (SM1) because of the similar environments (Fig. 4J). For $\boldsymbol{F}^0$, *a* appears longer than *c* in conformity with the resonance structures shown in Fig. 4A, where *c* is a double bond and not included in the conjugation pathway. For $\boldsymbol{F}^{-2}$, *a* and *c* appear with the same apparent length within the measurement accuracy, in agreement with the conjugation pathway highlighted in Fig. 4B including *a* and *c*. Interestingly, for $\boldsymbol{F}^{-1}$ bonds *a* and *c* have qualitatively the same relationship as in $\boldsymbol{F}^0$, i.e. *a* is larger than *c*, indicating that the conjugation pathway of the neutral molecule is maintained in the anion, even though a less stable 19π-electron system is formed. Another possible comparison is in each bond of the methine bridge, labelled $l_1$ and $l_2$ (Fig. 4K). These bonds significantly change for each charge state, but a simple picture of resonance Kekulé structures cannot rationalize their behavior. For both $\boldsymbol{F}^0$ and $\boldsymbol{F}^{-2}$, $l_1$ is shorter than $l_2$ with the difference between $l_1$ and $l_2$ being larger in $\boldsymbol{F}^{-2}$ than in $\boldsymbol{F}^0$. The latter indicates increased BLA for $\boldsymbol{F}^{-2}$ and thus likely a reduced aromaticity with respect to $\boldsymbol{F}^0$. For the anion $\boldsymbol{F}^{-1}$, $l_1$ is longer than $l_2$, i.e., the asymmetry at the meso-carbon position of the methine bridge is inverted with respect to both $\boldsymbol{F}^0$ and $\boldsymbol{F}^{-2}$. More insight is gained by DFT calculations of $\boldsymbol{F}$ adsorbed on NaCl in different charge states (Table S9-S11), in qualitative agreement with the experiment (Fig. S24). In porphine we observed significant changes of the bond-order relations upon charging and used them to learn about the evolution of the macrocycle's structure, aromaticity and conjugation pathway as a function of its oxidation state.

In conclusion, we demonstrated how structural, adsorption geometry and bond-order relations of individual molecules in multiple, controlled charge states can be resolved atomically by AFM. For porphine we observed significant structural differences indicating changes of conjugation pathway and aromaticity induced by charge-state transitions. Our method, coupled with redox potentials measurements (*10*) and orbital mapping (*33*) can provide a complete electrochemistry toolbox for single adsorbed molecules. Novel insights into fundamental processes where molecular ions play an important role, such as electron transfer, redox reactions, heterogeneous catalysis and organic photovoltaics can be obtained on the level of single molecules, even accessing doubly charged species.

**Acknowledgments:** We acknowledge Gerhard Meyer, Rolf Allenspach, Jascha Repp and Simon John Garden for discussions. **Funding:** The project was supported by the European Research Council Consolidator grant 'AMSEL' (Contract No. 682144). D. P. thanks Agencia Estatal de Investigación (MAT2016-78293-C6-3-R), Xunta de Galicia (Centro singular de investigación de Galicia, accreditation 2016–2019, ED431G/09), and the European Regional Development Fund for financial support. **Author contributions:** S.F. and L.G. designed the experiments. S.F. carried out the experiments with support from F.A.. S.F. and N.M. carried out the DFT calculations. All authors analyzed the data. S.F and L.G. drafted the manuscript and finalized it with the input from F.A., Y.Z., D. U., D. P. and N. M. **Competing interests:** Authors declare no competing interests. **Data and materials availability:** All data is available in the main text or the supplementary materials.

**Supplementary Materials:**

Materials and Methods

Supplementary Text SM1

Figures S1-S27

Tables S1-S11

References (*34-40*)

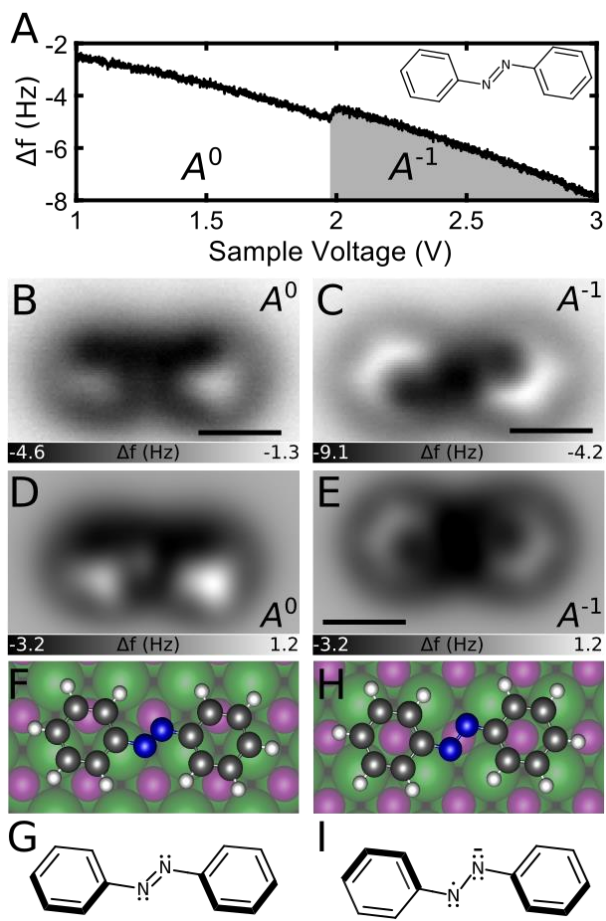

**Fig. 1. Measurements and calculations on azobenzene.** (A) $\Delta f(V)$ spectrum recorded on top of an azobenzene molecule. *V* was ramped from 1 to 3 V. The inset shows the chemical structure of azobenzene. (B) Constant-height AFM image of neutral azobenzene ($A^0$), at *V* = 0.5 V. (C) Constant-height AFM image of anionic azobenzene ($A^{-1}$), at *V* = 2.5 V, tip-sample distance reduced by 0.3 Å with respect to (B). (D, E) Simulated AFM images of on-surface $A^0$ and $A^{-1}$, respectively. All scale bars correspond to 5 Å. (F, H) Top view of the atomic models of $A^0$ and $A^{-1}$, respectively. (G, I) Chemical structures of $A^0$ and $A^{-1}$, respectively, with wedged bonds representing out-of-plane conformations.

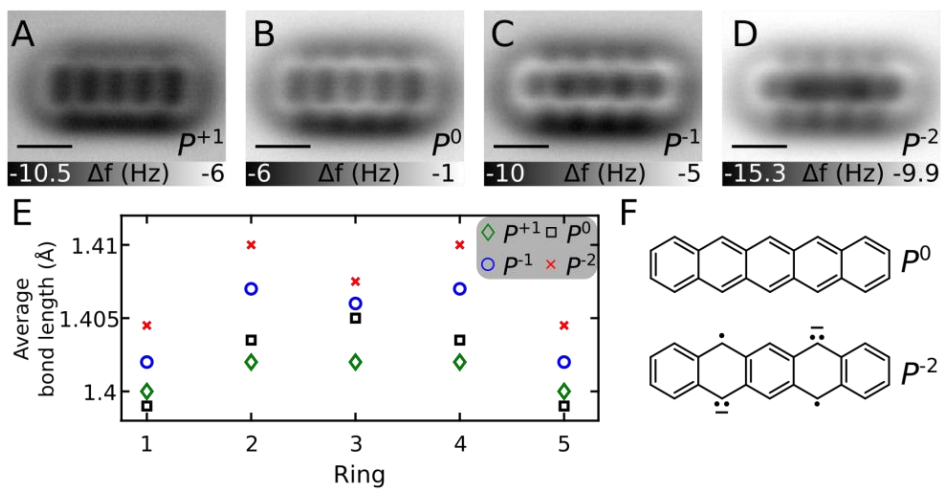

**Fig. 2. Measurements and calculations on pentacene.** Constant-height AFM images of the (A) cationic ($P^{+1}$), (B) neutral ($P^{0}$), (C) anionic ($P^{-1}$) and (D) dianionic ($P^{-2}$) molecule. Scale bars represent 5 Å. (B) and (C) are imaged at a tip-sample distance that was 0.3 Å smaller than in (A) and (D). $V$ is (A) -3.3 V, (B) 0.5 V, (C) 2.5 V and (D) 3.6 V. (E) DFT calculated average C-C bond length of each hydrocarbon ring for different charge states. (F) Possible resonance structures of $P^{0}$ and $P^{-2}$.

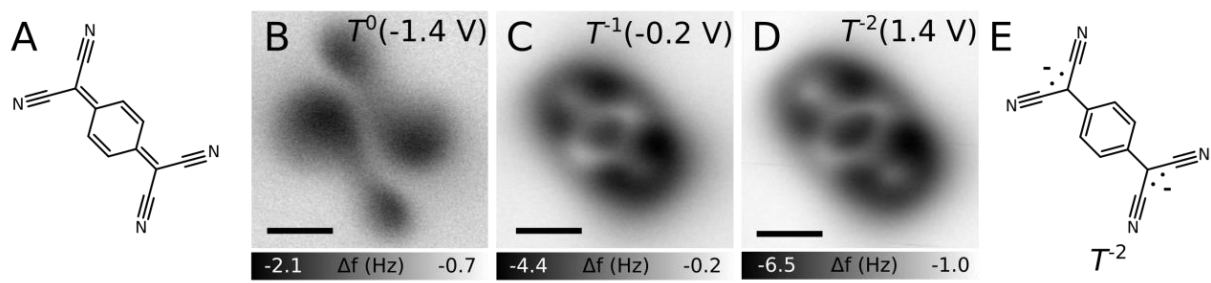

**Fig. 3. Tetracyanoquinodimethane model and measurements.** (A) Chemical structure of ***T***. Constant-height AFM images of the (B) neutral (***T⁰***), (C) anionic (***T⁻¹***) and (D) dianionic (***T⁻²***) molecule. (C) and (D) are imaged with a 1.9 Å smaller tip-sample distance than (B). Scale bars represent 5 Å. *V* is indicated in each image. (E) Major resonance structure proposed for ***T⁻²***.

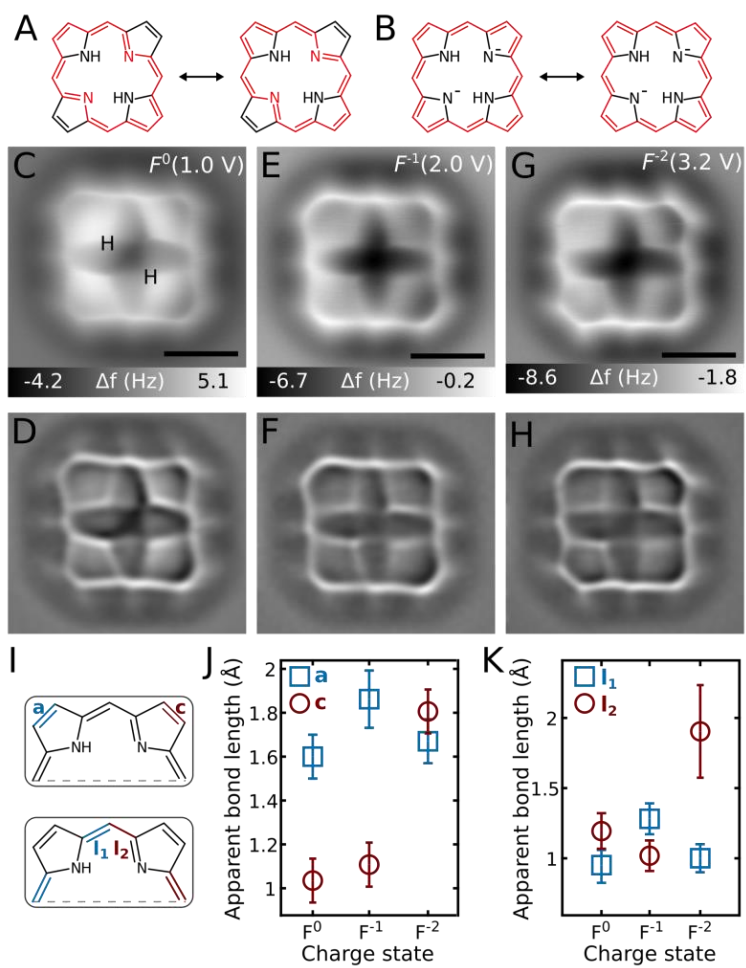

**Fig 4. Analysis of porphine and its conjugation pathway.** Chemical structure of (A) neutral ($F^0$) and (B) dianionic ($F^{-2}$) porphine. The red path shows the expected annulene-type conjugation pathway for each charge state. Constant-height and corresponding Laplace-filtered AFM images of (C, D) $F^0$, (E, F) $F^{-1}$ and (G, H) $F^{-2}$. The constant-height AFM images in (E) and (G) are taken at tip-sample distances larger by 0.5 Å and 0.4 Å, than the AFM image in (C), respectively. All scale bars correspond to 5 Å. $V$ is indicated in each image. (I) Highlighted bonds in $F$. Measured apparent bond lengths of (J) $a$ and $c$ and (K) $l_1$ and $l_2$ of $F$ as a function of charge state.

# Supplementary Materials for

## Molecular structure elucidation with charge-state control

Shadi Fatayer*, Florian Albrecht, Yunlong Zhang, Darius Urbonas, Diego Peña, Nikolaj Moll, and Leo Gross*.

Correspondence to: sfa@zurich.ibm.com, lgr@zurich.ibm.com

**This PDF file includes:**

**Materials and Methods**

By partially shielding the Cu(111) surface of a single crystal, we prepared a part of the surface with 2 monolayer (ML) NaCl islands and the other part with a 20 ML thick NaCl film. To that end, we evaporated NaCl with a nominal thickness of 20 ML onto the partially shielded sample and post annealed to 370 K, to form extended (001) oriented NaCl terraces. Afterwards, a nominal sub ML coverage of NaCl was evaporated on the whole surface at a sample temperature of 270 K to form 2 ML NaCl islands on the previously shielded side of the surface. The molecules and CO for tip functionalization were deposited onto the approximately 10 K cold sample. Prior to imaging molecules on the multilayer insulating film we prepared the tip on the area with 2ML NaCl islands, by indenting the tip into the Cu(111) surface and picking up CO from 2ML NaCl islands. The molecules were introduced into the UHV-system by being spread into a Si wafer. Afterwards, the wafer was flash-heated by resistive heating while facing the cold sample.

Frequency-shift ($\Delta f$) feedback was used to approach the tip on the thicker NaCl films. To this end the oscillation amplitude $A$ was increased to $A = 6$ Å to enhance the sensitivity to long-range forces (electrostatic force and van der Waals force). Once the tip had been approached to the sample, we used $\Delta f$ feedback for AFM imaging with typical $\Delta f$ setpoints on the order of -0.1 Hz to -1.0 Hz and a bias voltage of $V = 0.5$ V to locate molecules. Once a molecule was located, we turned to constant-height mode and the amplitude was decreased to $A = 0.5$ Å to increase sensitivity to short-range forces (Pauli repulsive force). For constant-height AFM images the tip was scanned on a plane parallel to the surface. The tip height $z$ was selected by applying a height offset with respect to the tip position with the $\Delta f$ feedback switched on or with respect to the preceeding measurement. For each shown set of AFM images, we define the tip height $z$ of the AFM image furthest away as $z = 0$ Å. The tip heights of the other AFM images within a set are given with respect to this tip height, with increasing $z$ corresponding to decreasing tip-sample distance. In general on hydrocarbon molecules, atomic contrast starts to occur for tip-sample distances near the minimum of $\Delta f(z)$ above the molecule, which corresponds to a distance of about 3.9 Å between the tip's oxygen atom and the imaged molecule's plane (*17*). All the $\Delta f(V)$ spectra were performed with $A = 0.5$ Å and at a tip-sample distance ranging from 0 Å to 1 Å away from the distance at which atomic contrast is achieved.

Density functional theory (DFT) was employed using the FHI-AIMS code (*34*). Each molecule, in different charge-state configurations, was independently investigated on a 4 layer thick NaCl cluster composed of 196 atoms. For each molecule the center position of the slab was changed to accommodate the appropriate adsorption position and orientation of the molecule. The geometries were optimized with the tight basis defaults. The Perdew-Burke-Ernzerhof exchange correlation functional was applied (*35*) with van der Waals correction (*36*) for structural relaxation. The convergence criterion was set at $10^{-3}$ eV/Å for the total forces and $10^{-5}$ eV for the total energies. The calculations were all spin-polarized.

**Supplementary text**

SM1. Changes of apparent bond length as a function of bias voltage V

AFM images are influenced by the tilting of the CO molecule at the tip apex (17, 24). In general, bonds often appear elongated due to overall (mainly van der Waals) attraction of the CO towards the molecule. Another effect of the CO tilting is the apparent sharpening of bonds, caused by the CO being repelled from regions of high electron density due to Pauli repulsion (17, 37). The lateral and vertical forces acting on the CO at the tip are changing with tip position and causing these distortions. Moreover, as has been shown, electrostatic forces also contribute to such distortions (37–39).

As discussed in the main text, we observed changes of apparent bond length as a function of applied bias voltage V. Such changes were observed for molecules in different charge states. In addition, we also observed voltage-dependent apparent bond length changes for a molecule in a fixed charge state. Pentacene in the neutral charge state can be measured over a wide range of applied bias voltages, from about V = -2.0 V to about V = +1.5 V, see Fig. S8, and therefore is investigated in detail to study this effect. We measured the apparent short and long axis of $P^0$ at different V for identical tip-sample distance, see Fig. S11. As shown in Fig. S11, both the apparent short and long axis shrink almost linearly with increasing V. The apparent short axis becomes shorter at a ratio of about -0.26 Å/V whereas this value is -0.11 Å/V for the long axis. The data suggests that the tilting of the CO at the tip apex is influenced by the electric field applied in the junction. The direction of the effect (apparent decrease in size with increasing V) indicates that with increasing V the CO is tilting less towards the molecule. This could be explained by (i) decreasing lateral forces tilting/attracting the CO less towards the molecule with increasing V, and/or (ii) an increase of the lateral spring constant of the CO tip due to increased attraction of CO in the vertical direction (towards the surface) with increasing V.

Figure S10 shows the apparent length of the molecular axes of the pentacene molecule in different charge states. As explained in the main text, different charge states are stabilized by applying a certain voltage, with increasing voltage resulting in the decrease of the charge state. We observe that the apparent short axis of pentacene decreases with negative charge (Fig. S10C). However, the long axis increases with negative charge (Fig. S10D). Qualitatively, DFT calculations also show the shrinkage in the short axis while the long axes elongates (Fig. S10 E, F). However, it is extremely challenging to deconvolve the effects from the electrostatic field, which are observed for $P^0$ (V dependent) versus effects from changed bond length in different charge states (V dependent and charge-state dependent) when comparing AFM images acquired at different voltages. Note that also the electric field in the junction will be modified when the molecule changes its charge state.

We observe qualitatively different trends for the evolution of the axes when looking at the effect of the voltage only in the case of $P^0$ (both short and long axis decrease with increasing V) and when comparing the different charge states at different voltages (short axis increases, long axis decreases with increasing V). This indicates that in the latter case, i.e. comparing images of different charge states at different voltages, both effects, i.e. the electrostatic field and the different bond length, contribute. Therefore, one must be extremely cautious when comparing absolute apparent bond length differences for AFM images of different charge states.

For this reason, as mentioned in the main text, we only compare individual bonds within an individual AFM image (obtained at a fixed voltage). Thus, we can obtain bond-order relations in a fixed charge state, i.e., we can resolve which bond is longer/shorter by comparing bonds in this charge state image that are in a similar chemical environment. Finally, we can compare bond-order relations obtained from images in different charge states.

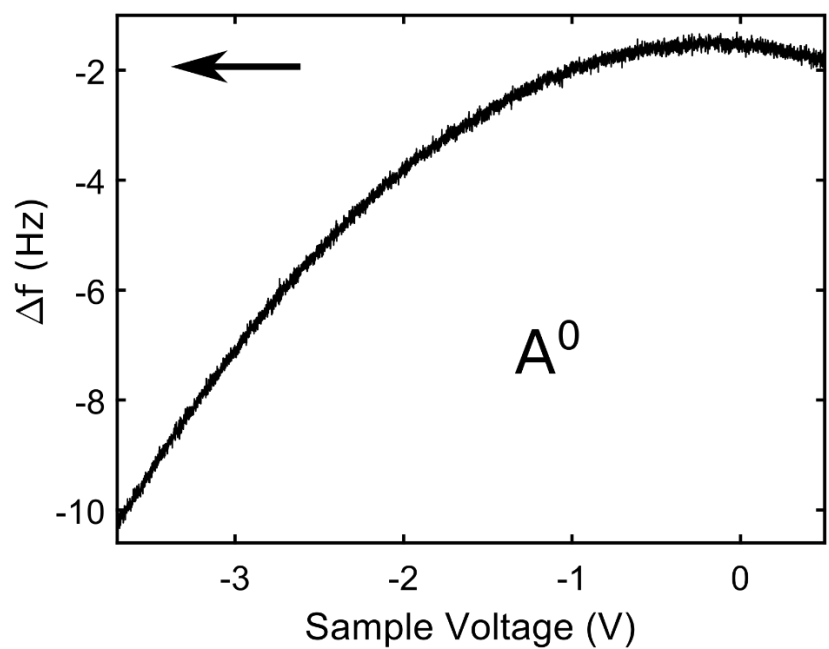

**Fig. S1. $\Delta f(V)$ spectrum of *A* for negative *V*.** Arrow indicates the direction of sample voltage ramp. No additional charge-state transitions are observed in the spectrum.

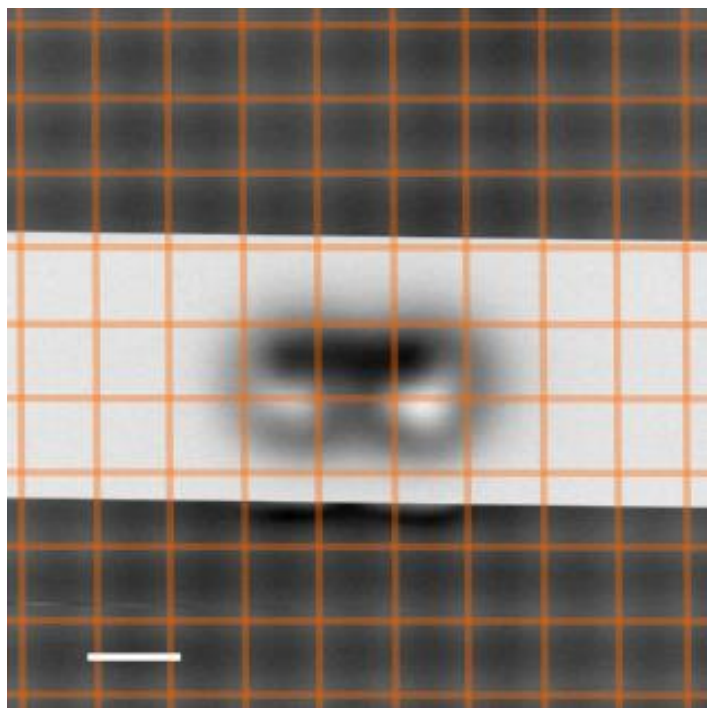

**Fig. S2. $A^0$ adsorption position.** Constant-height AFM image of the NaCl registry with the atomic resolution of $A^0$. Scale bar represents 5 Å. $V = 0.5$ V. In the middle part of the image the tip-sample distance was increased by 2.7 Å to atomically resolve azobenzene.

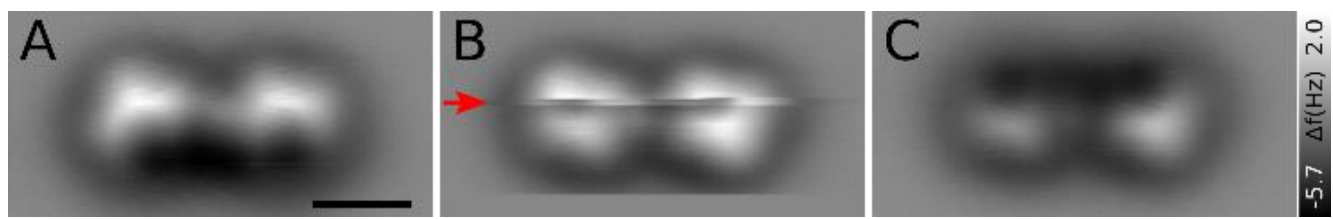

**Fig. S3. $A^0$ adsorption position manipulation.** Constant-height AFM images of $A^0$ before (A), while (B) and after (C) the manipulation. The red arrow in (B) indicates the manipulation event of $A^0$. Scale bar represents 5 Å. All images are recorded at $V = 0.5$ V.

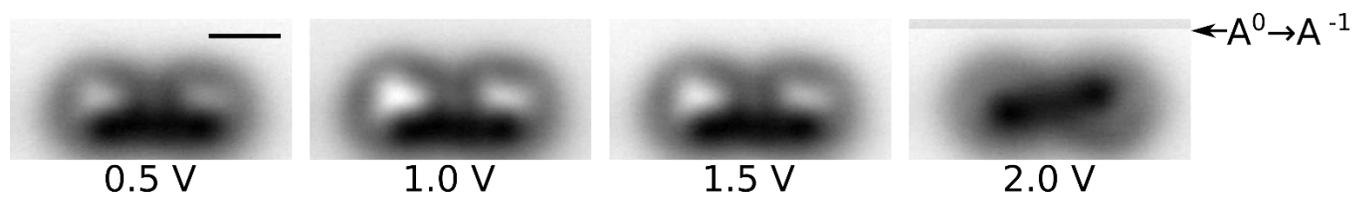

**Fig. S4. *V* dependent AFM imaging of *A*.** Constant-height AFM images performed at different *V*. Scale bar represents 5 Å.

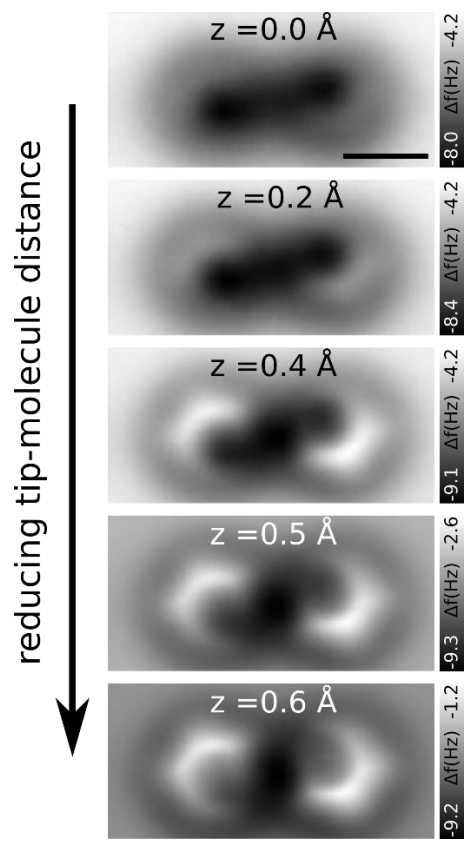

**Fig. S5. $A^{-1}$ constant-height AFM imaging as a function of tip height $z$.** Scale bars represent 5 Å.

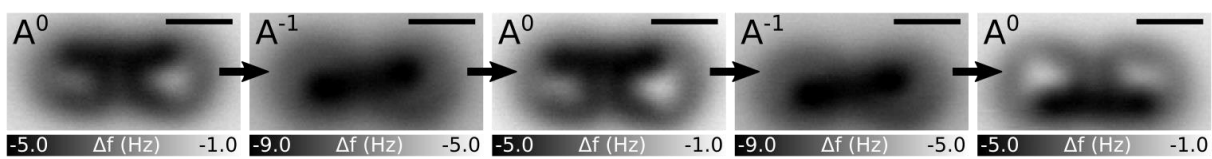

**Fig. S6. $A^0$ to $A^{-1}$ sequential switch.** Scale bars represents 5 Å. Images of $A^0$ have been recorded at $V = 0.5$ V while images of $A^{-1}$ have been recorded at $V = 2.5$ V.

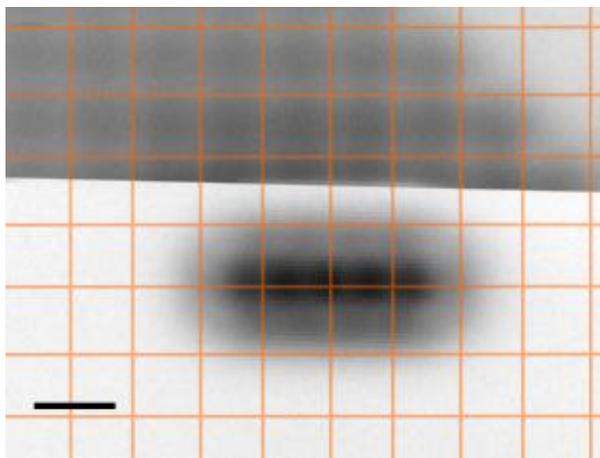

**Fig. S7. Adsorption position of *P*.** Constant-height AFM image of the NaCl registry with the atomic resolution of $P^0$. Scale bar represents 5 Å. $V = 0.5$ V. In the lower part of the image, the tip-sample distance was increased by 2.5 Å to resolve pentacene. The AFM images of *P* in different charge states showed that the molecule did not move and thus has the same adsorption site for all investigated charge states.

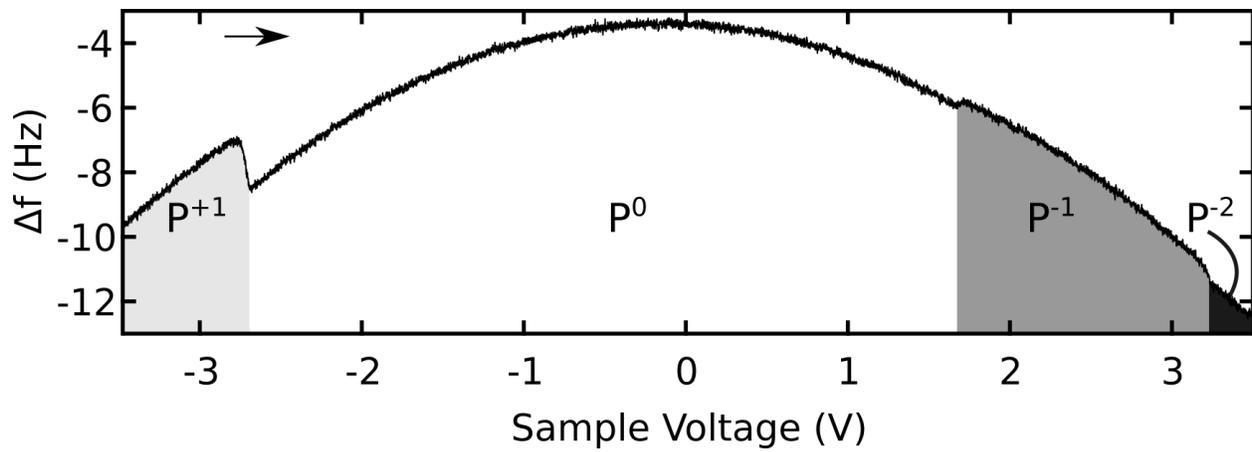

**Fig. S8. Δ*f*(*V*) spectrum of *P*.** Arrow indicates the direction of sample voltage ramp. Grey regions indicate the respective charge states.

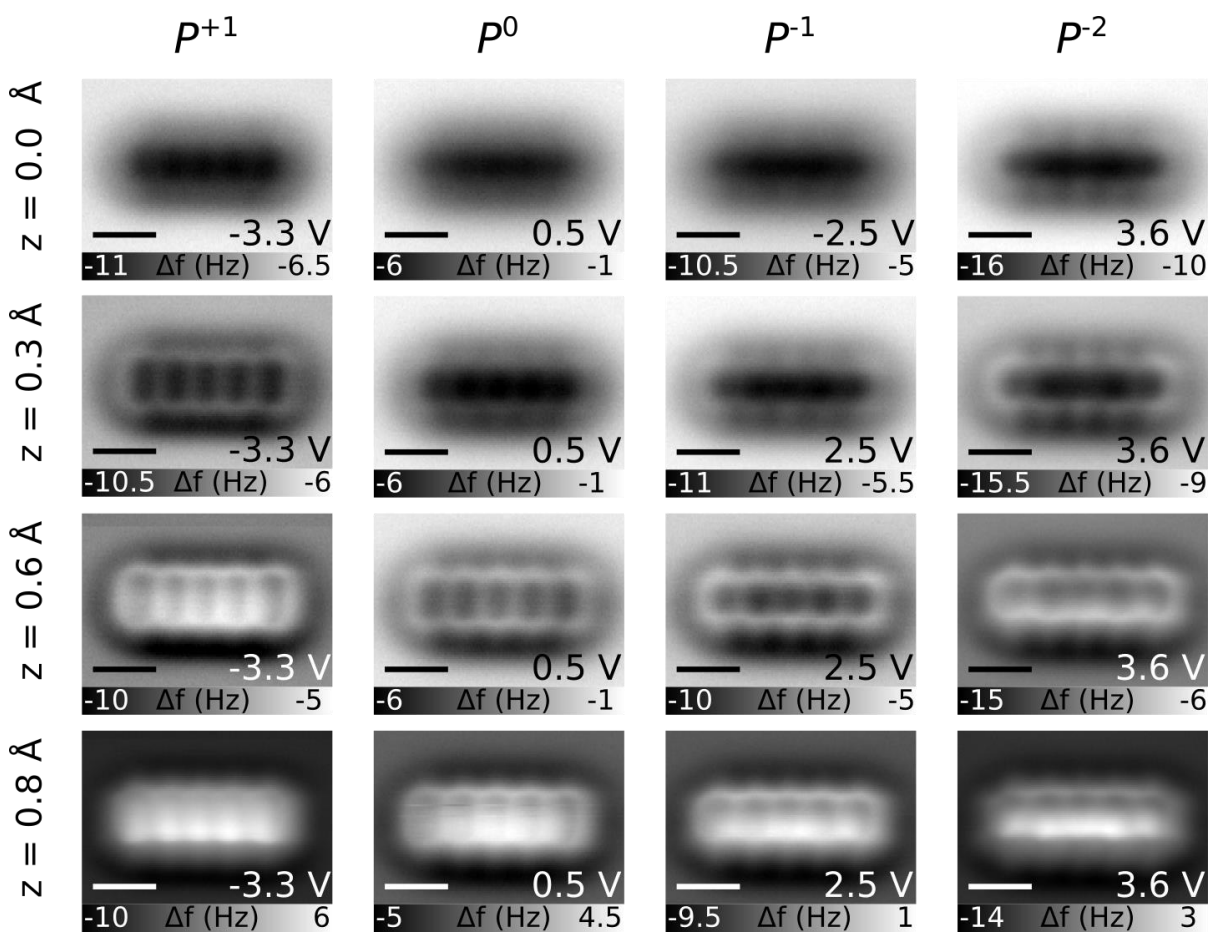

**Fig. S9. AFM data of *P* in different charge states and at different tip heights *z*.** $z = 0$ Å is assumed for the most far away image and increasing *z* equals decreasing tip-sample distance. All scale bars indicate 5 Å. *V* is indicated in each image.

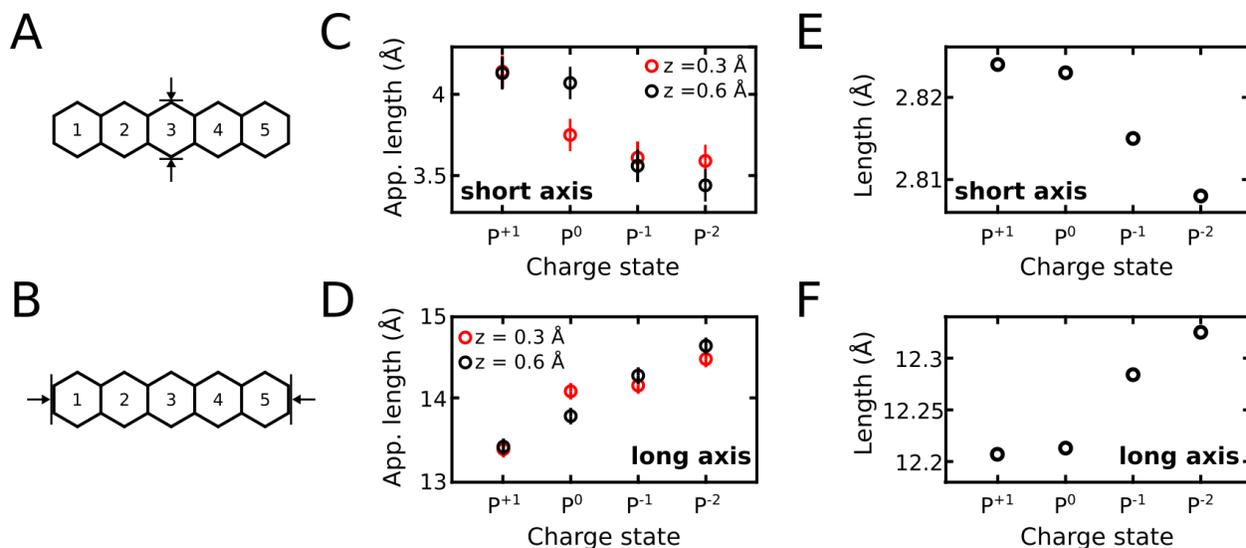

**Fig. S10. Measurements on the apparent long and short axis of *P* as a function of charge state.** Definition of the short (A) and long (B) axis of *P*. The apparent short axis is measured at the central ring. Measured short (C) and long (D) axis apparent length as a function of *V* for two different tip heights *z*. The *z* values are the same as in Fig. S9. *V* at which *P* was measured was -3.3 V (*P*$^{+1}$), 0.5 V (*P*$^0$), 2.5 V (*P*$^{-1}$) and 3.6 V (*P*$^{-2}$). DFT calculated lengths of the long (E) and short (F) axis of pentacene as a function of charge state.

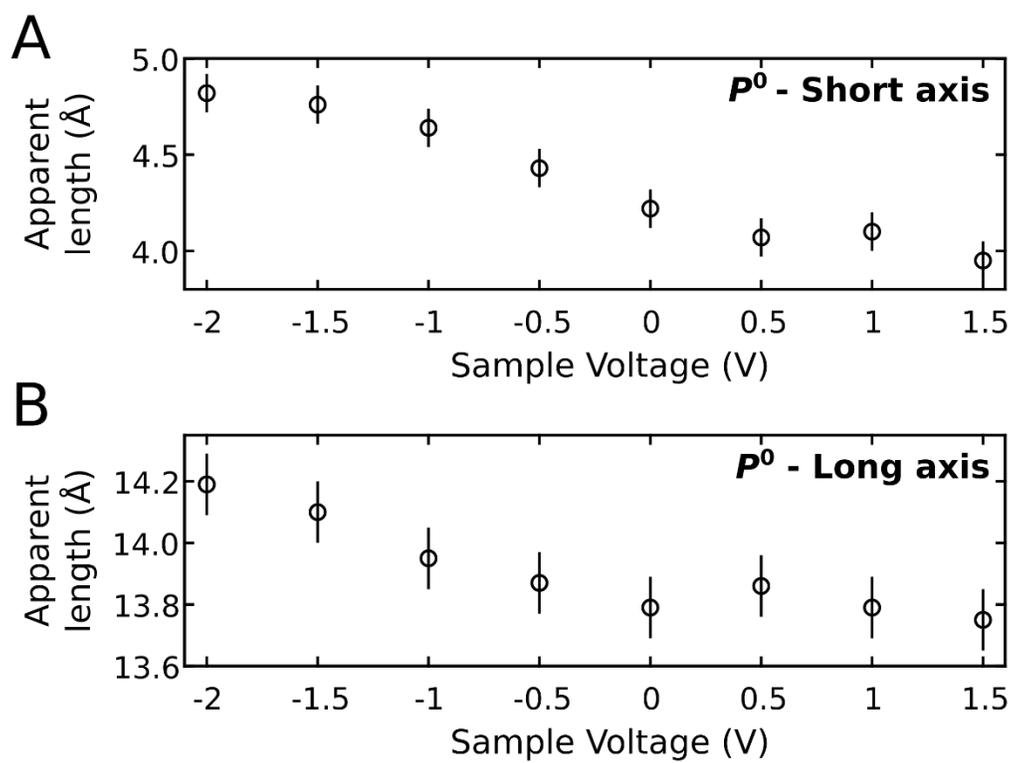

**Fig. S11. Measured short (A) and long (B) axis apparent length of $P^0$ as a function of *V*.** All measurements are performed at the same tip height.

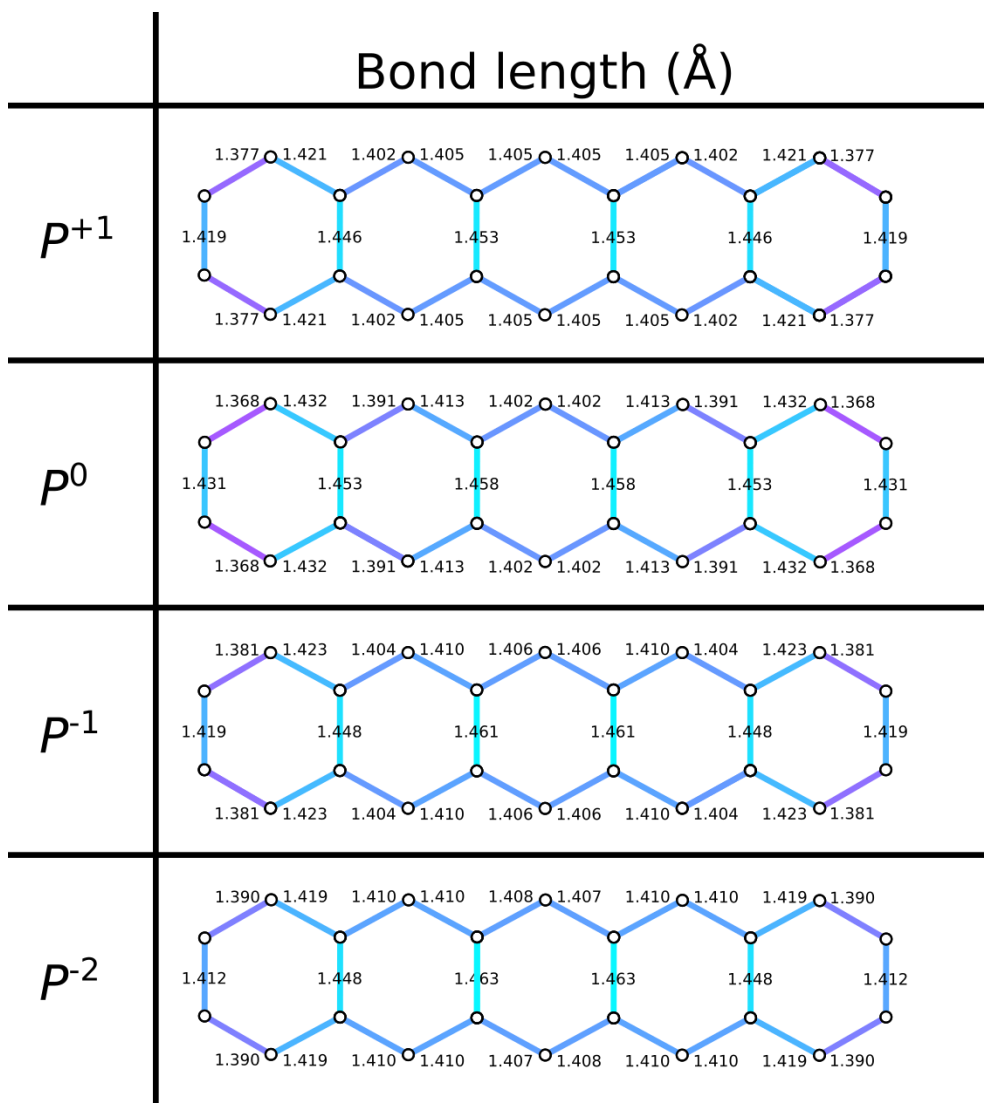

**Fig. S12. DFT obtained bond lengths of *P* for different charge states.**

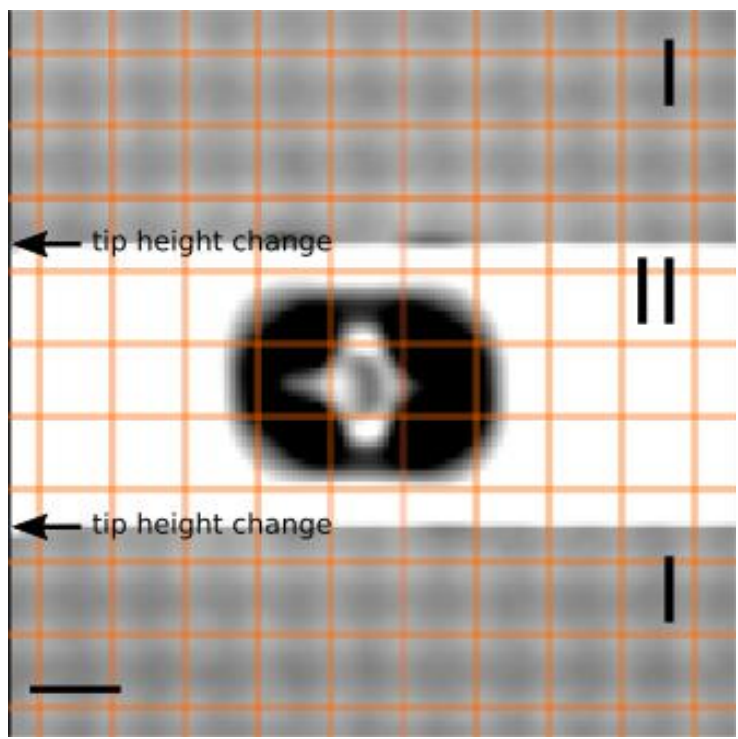

**Fig. S13. Adsorption position of $T^{-1}$.** Constant-height AFM image of the NaCl registry (region I) with the atomic resolution of $T^{-1}$ (region II). Scale bar represents 5 Å. $V = 0.0$ V. Tip-sample distance was increased by 1.9 Å in region II of the AFM image. The AFM images of $T$ in different charge states showed that the molecule did not move and thus has the same adsorption site for all investigated charge states.

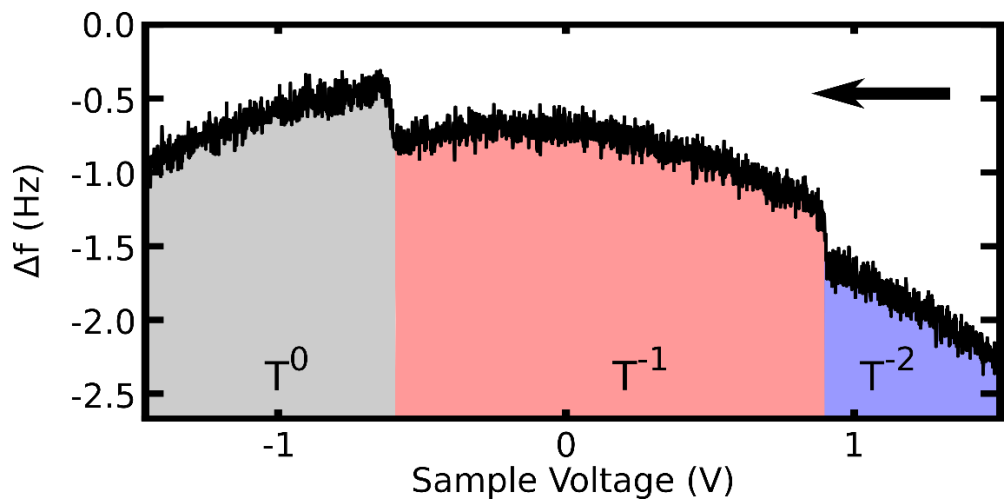

**Fig. S14. $\Delta f(V)$ spectrum of *T*.** Arrow indicates the direction of sample voltage ramp. Colored regions indicate the charge states.

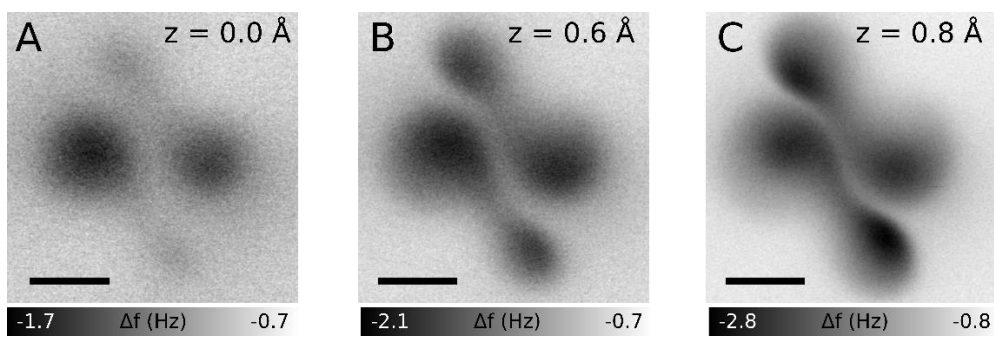

**Fig. S15. $T^0$ imaging as a function of tip height $z$.** Constant-height AFM images of $T^0$. Scale bar represents 5 Å. $V = -1.4$ V.

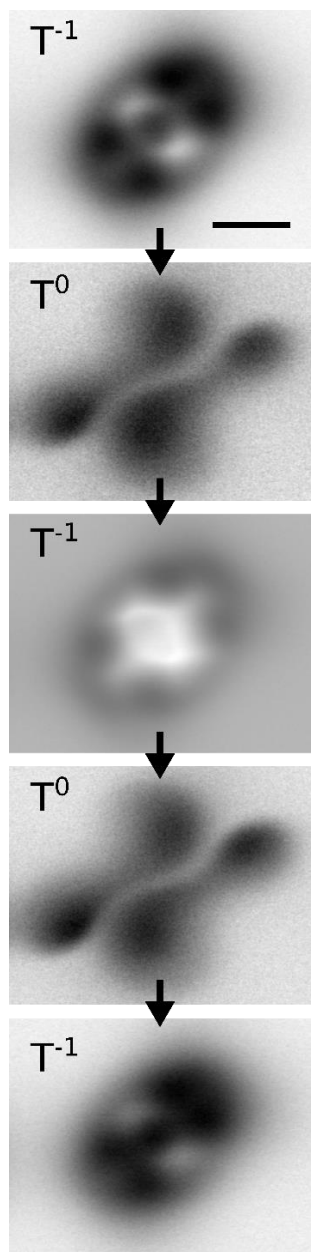

**Fig. S16.** $T^{-1}$ **to** $T^0$ **sequential switching.** Constant-height AFM images taken sequentially for $T$ in either the anionic or neutral state. AFM images of $T^{-1}$ and $T^0$ are taken at $V = 0.5$ V and $V = -1.4$ V, respectively. AFM images of $T^0$ correspond to $z = 0$ Å. The AFM images of $T^{-1}$ have been obtained at $z = 1.4$ Å, i.e. with the tip approaching the molecule by 1.4 Å with respect to the image in $T^0$. Scale bar represents 5 Å. Contrast changes in the images for the respective same charge state occurred because the $\Delta f$ was switched on between images, resulting in tip height offsets of a few 0.1 Å due to the $z$-noise in the $\Delta f$ feedback.

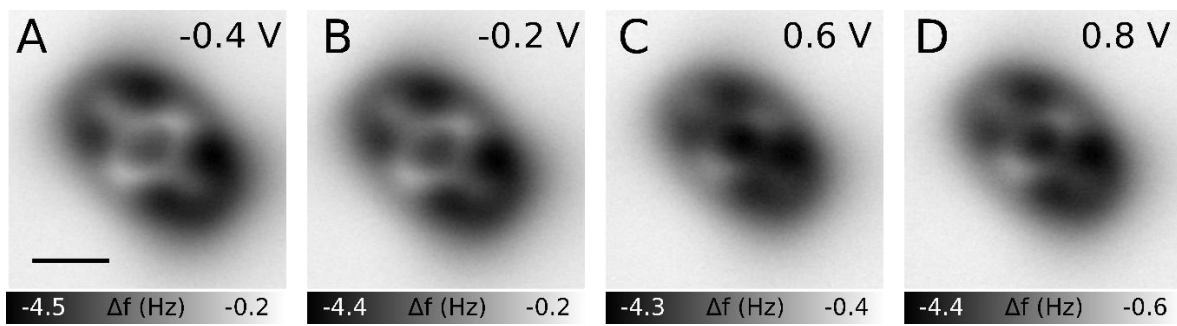

**Fig. S17. $T^{-1}$ imaging as a function of $V$.** Constant-height AFM images of $T^{-1}$. Scale bar represents 5 Å. $V$ is indicated in each image. Note that the contrast of the entire molecule with respect to the background changes, explained by different local contact potential differences (LCPDs) on the molecule and on NaCl, respectively. However, the relative brightness contrast within the molecule is qualitatively the same for all voltages, indicating the partial quinoidic character of the central ring of $T^{-1}$.

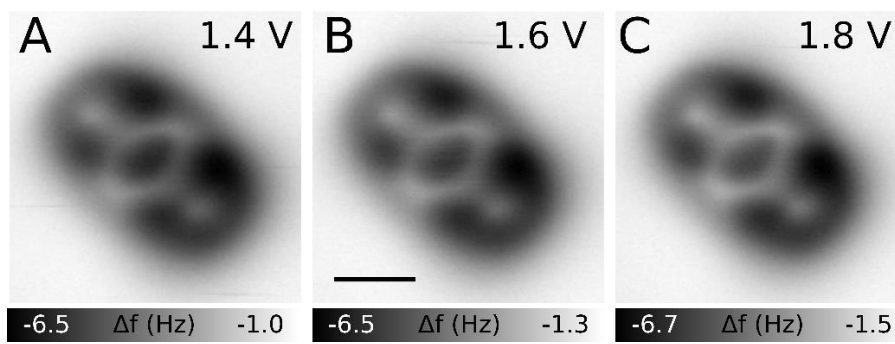

**Fig. S18.** $T^{-2}$ **imaging as a function of** $V$**.** Constant-height AFM images of $T^{-2}$. Scale bar represents 5 Å. $V$ is indicated in each image.

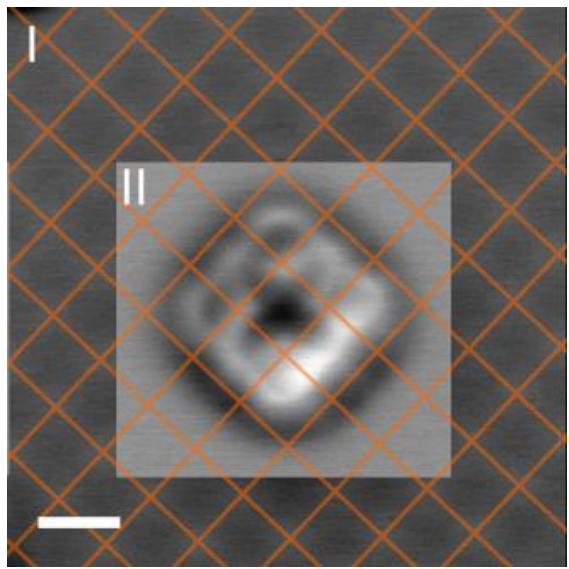

**Fig. S19. Adsorption position of *F*.** Constant-height AFM image of the NaCl registry (region I) with the atomic resolution of $F^0$ (region II). Scale bar represents 5 Å. $V$ = 0.0 V. Tip-sample distance was increased by 1.8 Å in region II of the AFM image.

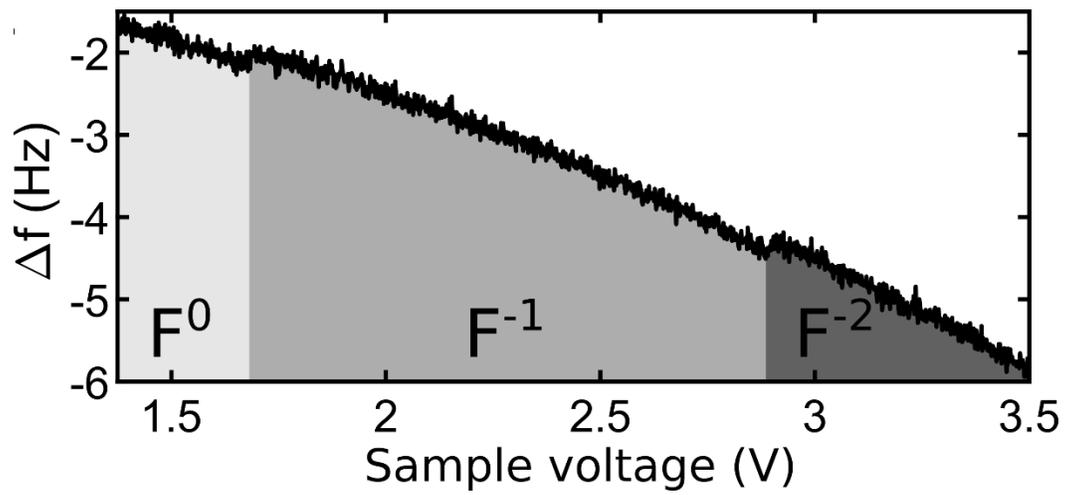

**Fig. S20.** $\Delta f(V)$ **spectrum of $F$.** Voltage ramp is towards larger sample voltages. Colored regions indicate the charge states.

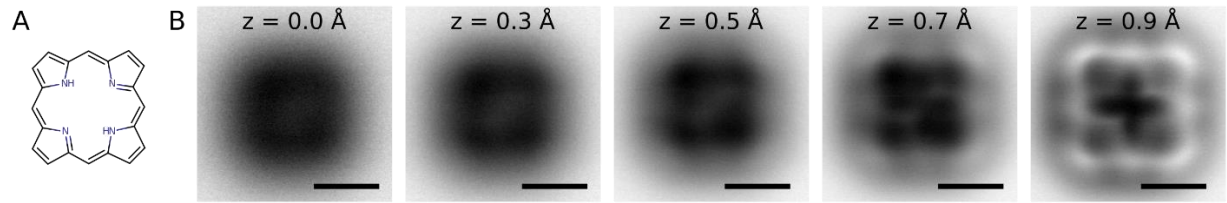

**Fig. S21. $F^0$ imaging as a function of tip height $z$.** (A) Chemical structure of **F**. (B) Constant-height AFM images of $F^0$. Scale bars indicates 5 Å. $V = 1.0$ V.

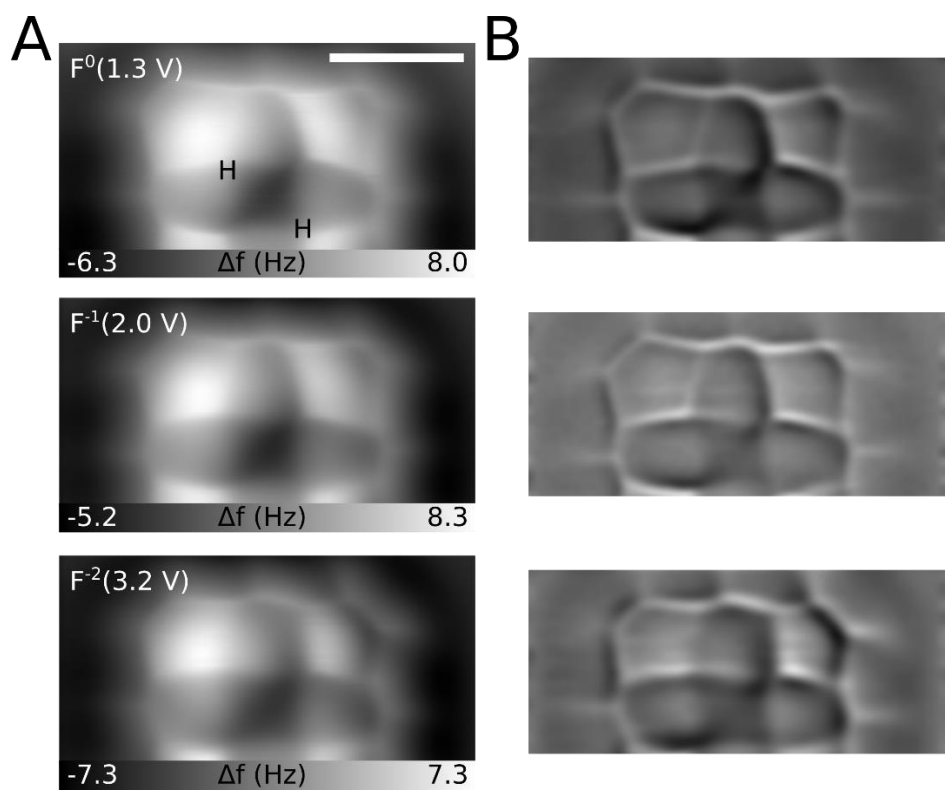

**Fig. S22. Zoomed-in AFM images of *F* in different charge states.** (A) Constant-height AFM images. Scale bars indicate 5 Å. *V* is indicated in each image. The positions of the hydrogen atoms in the cavity, as indicated in (A), are identical in the three measuremetns. All AFM images are taken at the same tip height. (B) Laplace-filtered respective AFM images of (A).

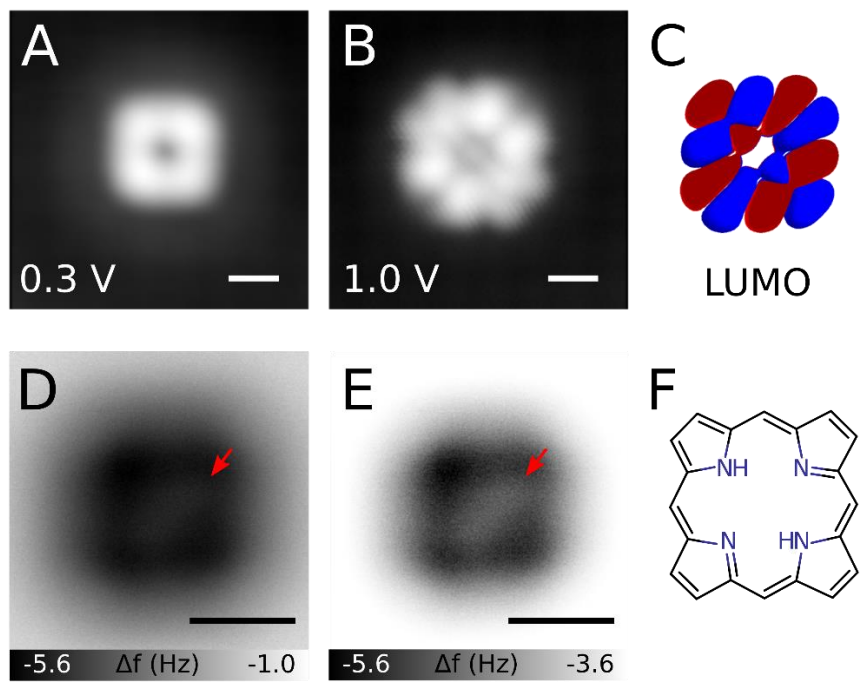

**Fig. S23. Analysis of *F* on 2 ML NaCl.** STM images of (A) in-gap and (B) at negative ion resonance of *F* using a CO functionalized tip. Tunneling current $I = 1$ pA for both images. *V* is indicated in each image. (C) DFT-calculated LUMO. (D) Constant-height AFM image of *F*. Tip-height offset is -0.4 Å with respect to the STM set point of $I = 1$ pA and $V = 0.2$ V. The red arrow indicates an inner contrast in the image, located inside the macrocycle of *F*. (E) Same AFM image as (D) but with an over-saturated scale. The red arrow indicates an inner contrast in the image. (F) Chemical structure of *F* based on the comparison of the experimental negative ion resonance in (B) with the DFT-calculated LUMO density. All scale bars indicate 5 Å. Similar AFM contrast related to the position of the hydrogens in the cavity was observed for the macrocycle in naphthalocyanine (*40*).

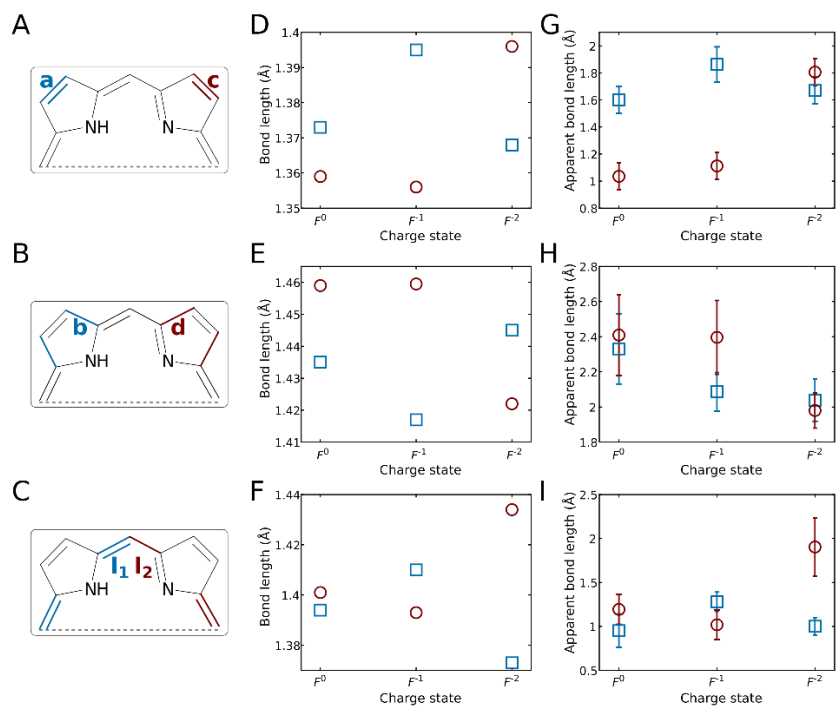

**Fig. S24. Theoretical and experimental bond length comparison.** (A to C) Half of the chemical structure of *F* with colors highlighting the investigated bonds. (D to F) Bond-length analysis of the DFT-calculated structures of *F* in different charge states. (G to I) Apparent bond-length analysis of the experimental AFM images of *F* in different charge states. (D and G) correspond to the bonds *a* and *c*. (E and H) correspond to the bonds *b* and *d*. (F and I) correspond to the bonds $l_1$ and $l_2$.

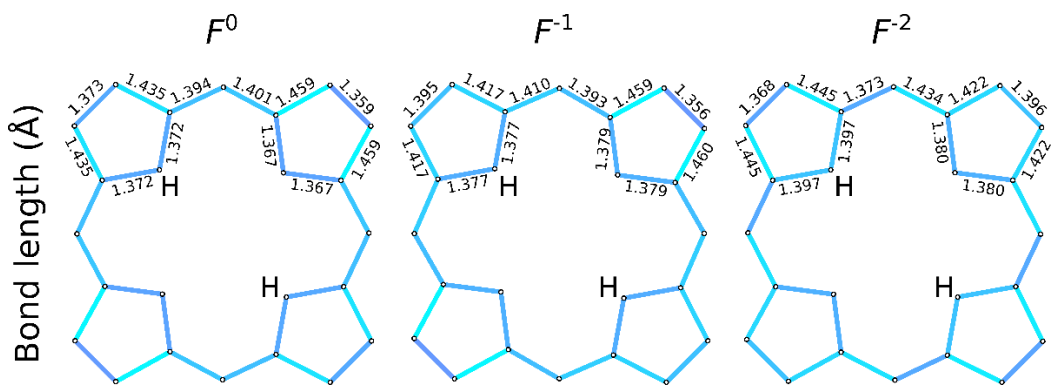

**Fig. S25. Bond lengths of *F* for different charge states.** Based on DFT calculations. Only the bond lengths in the upper part of *F* are shown. The bond lengths in the lower part of respective equivalent bonds are identical to the upper part.

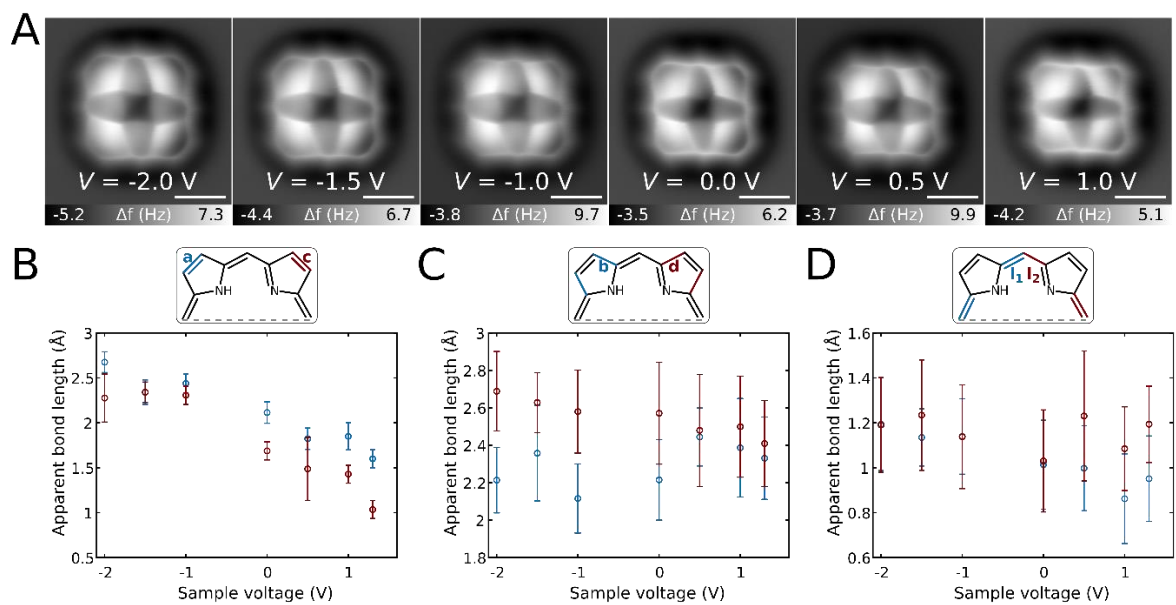

**Fig. S26. Analysis of $F^0$ as a function of $V$.** (A) Sequence of constant-height AFM images of $F^0$ at different $V$. $V$ is indicated in each image. (B-D) Apparent bond length analysis based on the AFM images in (A).

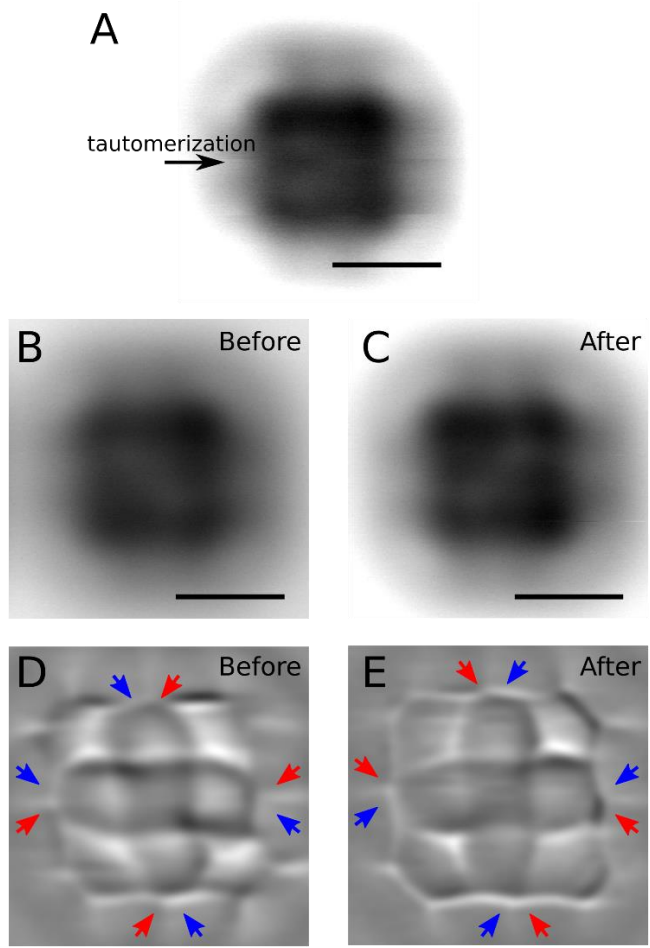

**Fig. S27. Tautomerization switching of *F*.** (A) Constant-height AFM image of *F*. The tautomerization switched, while imaging at $V = 3.7$ V, within the line indicated by the black arrow. Note that electrons are not exchanged between tip and molecule, as evidenced by the absence of charge-state switching. Thus the tautomerization switching in this experiment should be induced by the applied electric field. Constant-height AFM images before (B) and after (C) the tautomerization switch in (A). Scale bars indicate 5 Å. Laplace-filtered AFM images before (D) and after (E) the tautomerization switched. $z = 0.6$ Å and $V = 3.2$ V for both (D) and (E). Red and blue arrows indicate apparent short and long bonds in the methine bridge of $F^{-2}$, respectively. No tautomerization switching was observed for $V < 3.7$ V.

**Table S1. DFT calculated geometry of $A^0$ on 4 ML NaCl. All values reported are in Å.**

| | | | |
|---|---:|---:|---:|
| N |  0.69393881 |  1.26888351 | 3.56421910 |
| N |  0.73938175 |  0.18498877 | 2.91348442 |
| C | -2.81478753 |  2.35472512 | 2.91116959 |
| C | -2.67627174 |  3.72822161 | 3.13688832 |
| C | -1.43773187 |  4.25488201 | 3.51664357 |
| C | -0.34214362 |  3.41351336 | 3.66723490 |
| C | -0.46848372 |  2.04191133 | 3.39378571 |
| C | -1.71619277 |  1.50956664 | 3.03082980 |
| C |  1.90049751 | -0.59474471 | 3.16244078 |
| C |  1.79790945 | -1.96425137 | 2.88434050 |
| C |  2.91060834 | -2.79031972 | 3.02909735 |
| C |  4.13244145 | -2.24838640 | 3.43596935 |
| C |  4.23051336 | -0.88250775 | 3.72272593 |
| C |  3.12181085 | -0.05432308 | 3.59404267 |
| H | -3.78614900 |  1.94761024 | 2.62841083 |
| H | -1.81575665 |  0.43987199 | 2.85933694 |
| H |  0.63379778 |  3.79576792 | 3.96566165 |
| H | -1.33299086 |  5.32481930 | 3.69505053 |
| H | -3.53573862 |  4.38591680 | 3.01906300 |
| H |  0.84500530 | -2.36392190 | 2.54052817 |
| H |  5.01253239 | -2.88564374 | 3.50944453 |
| H |  2.82949748 | -3.85403758 | 2.80043920 |
| H |  3.18745288 |  1.01533151 | 3.78374786 |
| H |  5.18752682 | -0.45946018 | 4.02949694 |

**Table S2. DFT calculated geometry of *A*$^{-1}$ on 4 ML NaCl. All values reported are in Å.**

| | | | |
|---|---:|---:|---:|
| N |  0.10537002 |  0.66078701 | 3.10284173 |
| N | -0.10567941 | -0.66148571 | 3.10166772 |
| C | -3.45698814 |  1.78392001 | 3.37190600 |
| C | -3.27914145 |  3.17503188 | 3.35979030 |
| C | -1.97190067 |  3.68625628 | 3.28313992 |
| C | -0.88178618 |  2.83066686 | 3.20315603 |
| C | -1.04469215 |  1.41983008 | 3.20114577 |
| C | -2.37247703 |  0.91651919 | 3.30271258 |
| C |  1.04429584 | -1.42057868 | 3.20093930 |
| C |  0.88145011 | -2.83142855 | 3.20354334 |
| C |  1.97163000 | -3.68691608 | 3.28376432 |
| C |  3.27889238 | -3.17562360 | 3.35989203 |
| C |  3.45678776 | -1.78457415 | 3.37179368 |
| C |  2.37206697 | -0.91714611 | 3.30271109 |
| H | -4.46743440 |  1.37427360 | 3.42424638 |
| H | -2.51591718 | -0.16099165 | 3.30294680 |
| H |  0.13207423 |  3.22169215 | 3.12445045 |
| H | -1.81471846 |  4.76494002 | 3.26330125 |
| H | -4.13721680 |  3.84456949 | 3.39945366 |
| H | -0.13239816 | -3.22252306 | 3.12515414 |
| H |  4.13694160 | -3.84510400 | 3.39960749 |
| H |  1.81444741 | -4.76556518 | 3.26422065 |
| H |  2.51551319 |  0.16040212 | 3.30253041 |
| H |  4.46712474 | -1.37470080 | 3.42387497 |

**Table S3. DFT calculated geometry of $P^0$ on 4 ML NaCl. All values reported are in Å.**

| | | | |
|---|---:|---:|---:|
| C | 0.75071788 | -2.73613067 | 3.12299414 |
| C | 2.09081735 | -3.10716750 | 3.10593385 |
| C | 3.11624420 | -2.07796916 | 3.10315993 |
| C | 2.74035958 | -0.73918985 | 3.11874589 |
| C | 1.38309504 | -0.34766839 | 3.13907340 |
| C | 0.35422940 | -1.38027925 | 3.14051183 |
| C | -0.99481789 | -0.99989681 | 3.14198603 |
| C | -1.38007755 | 0.34775161 | 3.13962920 |
| C | -0.35118856 | 1.38033660 | 3.13975982 |
| C | 0.99783802 | 0.99996698 | 3.14073963 |
| C | 4.48945829 | -2.47961013 | 3.04998702 |
| C | 4.83202166 | -3.80347383 | 2.99846982 |
| C | 3.82231870 | -4.81692985 | 3.00191356 |
| C | 2.49739633 | -4.47903608 | 3.05620189 |
| C | -2.73733759 | 0.73935570 | 3.11978559 |
| C | -3.11316304 | 2.07815861 | 3.10342482 |
| C | -2.08767383 | 3.10730481 | 3.10485189 |
| C | -0.74759798 | 2.73619870 | 3.12144886 |
| C | -2.49417605 | 4.47917682 | 3.05431531 |
| C | -3.81910239 | 4.81711678 | 3.00048814 |
| C | -4.82887854 | 3.80371851 | 2.99837720 |
| C | -4.48637747 | 2.47987407 | 3.05071822 |
| H | -5.87614554 | 4.09969966 | 2.94728034 |
| H | -5.25300523 | 1.70320179 | 3.04108072 |
| H | -1.72020235 | 5.24859639 | 3.04778085 |
| H | -4.11881955 | 5.86335291 | 2.95102690 |
| H | -3.51090125 | -0.03206244 | 3.10813406 |
| H | 0.02107477 | 3.51250267 | 3.11098606 |
| H | -1.76509070 | -1.77292186 | 3.12120624 |
| H | 1.76808637 | 1.77297892 | 3.11900693 |
| H | -0.01789673 | -3.51250472 | 3.11350341 |
| H | 3.51386694 | 0.03226187 | 3.10608046 |
| H | 4.12207713 | -5.86317574 | 2.95304960 |
| H | 1.72349695 | -5.24854389 | 3.05070565 |
| H | 5.25600797 | -1.70287645 | 3.03932926 |
| H | 5.87928446 | -4.09939877 | 2.94697154 |

**Table S4. DFT calculated geometry of *P*[-1] on 4 ML NaCl. All values reported are in Å.**

| | | | |
|---|---:|---:|---:|
| C | 0.75381995 | -2.74033232 | 3.00943096 |
| C | 2.10356440 | -3.12730030 | 3.01366221 |
| C | 3.12724742 | -2.10361630 | 3.01373903 |
| C | 2.74024271 | -0.75386111 | 3.00959657 |
| C | 1.38802144 | -0.35498532 | 3.02183394 |
| C | 0.35489615 | -1.38812349 | 3.02174770 |
| C | -0.99523602 | -0.99537399 | 3.02977034 |
| C | -1.38803996 | 0.35476879 | 3.02174296 |
| C | -0.35492029 | 1.38792918 | 3.02180370 |
| C | 0.99522299 | 0.99519130 | 3.02988896 |
| C | 4.49187915 | -2.50693757 | 2.98193100 |
| C | 4.84452488 | -3.84139865 | 2.94263530 |
| C | 3.84142675 | -4.84452545 | 2.94261286 |
| C | 2.50692984 | -4.49192089 | 2.98177948 |
| C | -2.74023416 | 0.75367200 | 3.00950001 |
| C | -3.12723594 | 2.10341941 | 3.01371647 |
| C | -2.10360174 | 3.12711504 | 3.01373845 |
| C | -0.75383669 | 2.74016090 | 3.00950920 |
| C | -2.50692763 | 4.49176042 | 2.98179487 |
| C | -3.84141171 | 4.84437122 | 2.94252386 |
| C | -4.84451182 | 3.84121279 | 2.94255021 |
| C | -4.49189289 | 2.50673926 | 2.98181010 |
| H | -5.89479932 | 4.13119612 | 2.90929625 |
| H | -5.26111976 | 1.73097492 | 2.98058871 |
| H | -1.73119816 | 5.26102669 | 2.98057056 |
| H | -4.13140388 | 5.89465763 | 2.90924347 |
| H | -3.51683693 | -0.01651982 | 3.01465459 |
| H | 0.01636851 | 3.51676780 | 3.01475407 |
| H | -1.76754233 | -1.76766532 | 3.02839923 |
| H | 1.76751776 | 1.76751831 | 3.02860681 |
| H | -0.01634707 | -3.51696931 | 3.01459974 |
| H | 3.51683950 | 0.01635407 | 3.01493184 |
| H | 4.13139654 | -5.89480852 | 2.90935905 |
| H | 1.73118320 | -5.26119423 | 2.98048817 |
| H | 5.26111281 | -1.73116552 | 2.98075866 |
| H | 5.89481117 | -4.13137973 | 2.90950229 |

**Table S5. DFT calculated geometry of $P^{+1}$ on 4 ML NaCl. All values reported are in Å.**

| | | | |
|---|---|---|---|
| C | 0.74491781 | -2.73106653 | 3.23140575 |
| C | 2.09515574 | -3.10717734 | 3.22991849 |
| C | 3.11558457 | -2.08320885 | 3.22790659 |
| C | 2.73485046 | -0.73428495 | 3.22844710 |
| C | 1.38319748 | -0.35162237 | 3.23492573 |
| C | 0.35761418 | -1.38074729 | 3.23592116 |
| C | -0.99507914 | -1.00019924 | 3.22301918 |
| C | -1.38023920 | 0.35117378 | 3.23527413 |
| C | -0.35465912 | 1.38029136 | 3.23544175 |
| C | 0.99803156 | 0.99974581 | 3.22222492 |
| C | 4.48149785 | -2.47324265 | 3.18803421 |
| C | 4.82550089 | -3.80623366 | 3.14856657 |
| C | 3.82397004 | -4.81126264 | 3.15107654 |
| C | 2.48988015 | -4.47177626 | 3.19262893 |
| C | -2.73188690 | 0.73383431 | 3.22905946 |
| C | -3.11261554 | 2.08275208 | 3.22796175 |
| C | -2.09218048 | 3.10671671 | 3.22912209 |
| C | -0.74194934 | 2.73060764 | 3.23037554 |
| C | -2.48690265 | 4.47130124 | 3.19127641 |
| C | -3.82100289 | 4.81076772 | 3.15002255 |
| C | -4.82254153 | 3.80574458 | 3.14835031 |
| C | -4.47853785 | 2.47277477 | 3.18835082 |
| H | -5.86984491 | 4.09985327 | 3.09971561 |
| H | -5.24528932 | 1.69757645 | 3.16828714 |
| H | -1.71437924 | 5.24083389 | 3.17367734 |
| H | -4.11874201 | 5.85706951 | 3.10268964 |
| H | -3.50528220 | -0.03654850 | 3.20423776 |
| H | 0.02576363 | 3.50668466 | 3.20647907 |
| H | -1.76507974 | -1.77277426 | 3.18456814 |
| H | 1.76800451 | 1.77231270 | 3.18313081 |
| H | -0.02279630 | -3.50716215 | 3.20816492 |
| H | 3.50824848 | 0.03607198 | 3.20295857 |
| H | 4.12168752 | -5.85758673 | 3.10416921 |
| H | 1.71737238 | -5.24134529 | 3.17575074 |
| H | 5.24824094 | -1.69805112 | 3.16735595 |
| H | 5.87279590 | -4.10033803 | 3.09973731 |

**Table S6. DFT calculated geometry of *P*[-2] on 4 ML NaCl. All values reported are in Å.**

| | | | |
|---|---:|---:|---:|
| C |  0.76575198 | -2.74120426 | 3.05520579 |
| C |  2.11960929 | -3.13413678 | 3.05004888 |
| C |  3.14189680 | -2.10869222 | 3.04873588 |
| C |  2.74480954 | -0.75597903 | 3.05308975 |
| C |  1.39322301 | -0.35500867 | 3.06433016 |
| C |  0.36061477 | -1.39080222 | 3.06509909 |
| C | -0.98988662 | -0.99441256 | 3.06766134 |
| C | -1.39046990 |  0.35489011 | 3.06454562 |
| C | -0.35785406 |  1.39067634 | 3.06473006 |
| C |  0.99264485 |  0.99428786 | 3.06709523 |
| C |  4.50268353 | -2.51025396 | 3.00887578 |
| C |  4.86397624 | -3.85107286 | 2.95833388 |
| C |  3.86720635 | -4.85092986 | 2.95986663 |
| C |  2.52538473 | -4.49376725 | 3.01162829 |
| C | -2.74206439 |  0.75589850 | 3.05350171 |
| C | -3.13914943 |  2.10863824 | 3.04884336 |
| C | -2.11683643 |  3.13406794 | 3.04958341 |
| C | -0.76297324 |  2.74109470 | 3.05451388 |
| C | -2.52259150 |  4.49371020 | 3.01085017 |
| C | -3.86441825 |  4.85089477 | 2.95922523 |
| C | -4.86122119 |  3.85105075 | 2.95825015 |
| C | -4.49994756 |  2.51023675 | 3.00918417 |
| H | -5.91386402 |  4.13520730 | 2.91963450 |
| H | -5.26930255 |  1.73315755 | 3.01225847 |
| H | -1.74784863 |  5.26546119 | 3.01571198 |
| H | -4.15189376 |  5.90261986 | 2.92133870 |
| H | -3.51814687 | -0.01631466 | 3.07103720 |
| H |  0.00687529 |  3.51951624 | 3.07272103 |
| H | -1.76195694 | -1.76881032 | 3.07377987 |
| H |  1.76470619 |  1.76869289 | 3.07274043 |
| H | -0.00406585 | -3.51964265 | 3.07386792 |
| H |  3.52088291 |  0.01624662 | 3.07007865 |
| H |  4.15470419 | -5.90265021 | 2.92227009 |
| H |  1.75066522 | -5.26554248 | 3.01698741 |
| H |  5.27201909 | -1.73315639 | 3.01146260 |
| H |  5.91662143 | -4.13518403 | 2.91953614 |

**Table S7. DFT calculated geometry of *T*<sup>-1</sup> on 4 ML NaCl. All values reported are in Å.**

| | | | |
|---|---|---|---|
| N | 4.41489102 | -1.16311953 | 2.61815185 |
| N | 1.15821362 | -4.42372478 | 2.62004119 |
| N | -1.16357311 | 4.41465049 | 2.61506348 |
| N | -4.42170456 | 1.15578805 | 2.61895945 |
| C | 3.35553222 | -1.62417415 | 2.81427423 |
| C | 1.61700537 | -3.36335936 | 2.81594454 |
| C | -1.62330088 | 3.35511637 | 2.81290442 |
| C | -3.36201883 | 1.61605641 | 2.81495399 |
| C | -2.02017395 | 2.01343383 | 2.97608259 |
| C | 2.01362455 | -2.02130602 | 2.97567399 |
| C | 0.37026438 | 1.34206185 | 3.16415215 |
| C | 1.34278003 | 0.36916658 | 3.16416627 |
| C | 1.00228629 | -1.00956633 | 3.12551928 |
| C | -0.37655396 | -1.34957216 | 3.16600832 |
| C | -1.34900374 | -0.37672491 | 3.16660555 |
| C | -1.00852888 | 1.00195961 | 3.12613840 |
| H | -0.66320217 | -2.39968623 | 3.15122566 |
| H | -2.39934659 | -0.66251249 | 3.15217500 |
| H | 0.65661770 | 2.39222470 | 3.14801625 |
| H | 2.39305908 | 0.65509283 | 3.14831058 |

**Table S8. DFT calculated geometry of *T*$^{-2}$ on 4 ML NaCl. All values reported are in Å.**

| | | | |
|---|---:|---:|---:|
| N |  4.44987225 | -1.16693195 | 2.69204072 |
| N |  1.16734443 | -4.44966619 | 2.69237296 |
| N | -1.16686908 |  4.45027159 | 2.69298471 |
| N | -4.44945090 |  1.16766149 | 2.69264298 |
| C |  3.38380152 | -1.64762448 | 2.83449337 |
| C |  1.64796519 | -3.38353620 | 2.83470223 |
| C | -1.64744637 |  3.38414428 | 2.83508377 |
| C | -3.38329197 |  1.64828223 | 2.83478201 |
| C | -2.05004754 |  2.05091069 | 2.94513671 |
| C |  2.05061799 | -2.05029063 | 2.94480716 |
| C |  0.35711973 |  1.34044072 | 3.11997222 |
| C |  1.34048514 |  0.35695423 | 3.12005728 |
| C |  1.01374853 | -1.01330959 | 3.08394940 |
| C | -0.35660387 | -1.33991257 | 3.11987828 |
| C | -1.33993182 | -0.35643104 | 3.11990558 |
| C | -1.01314520 |  1.01385729 | 3.08401912 |
| H | -0.65282993 | -2.38809207 | 3.10395683 |
| H | -2.38812562 | -0.65249481 | 3.10407496 |
| H |  0.65327522 |  2.38862862 | 3.10439699 |
| H |  2.38871609 |  0.65302593 | 3.10437204 |

**Table S9. DFT calculated geometry of $F^0$ on 4 ML NaCl. All values reported are in Å.**

| | | | |
|---|---:|---:|---:|
| N | -2.11365370 | 0.00060052 | 3.22561536 |
| N | 0.00031843 | -2.01496261 | 3.18567918 |
| N | 2.11376640 | 0.00005226 | 3.22552123 |
| N | -0.00015361 | 2.01562583 | 3.18700223 |
| C | -4.25980264 | 0.68692652 | 3.10037566 |
| C | -4.25964665 | -0.68616174 | 3.10035661 |
| C | -2.89490687 | -1.12557024 | 3.16306833 |
| C | -2.89516573 | 1.12660586 | 3.16316824 |
| C | -2.42371175 | 2.43798867 | 3.12403189 |
| C | -2.42335611 | -2.43690745 | 3.12355419 |
| C | -1.08376206 | -2.84619995 | 3.13606913 |
| C | 1.08423776 | -2.84643106 | 3.13584171 |
| C | 0.67968984 | -4.24603112 | 3.06670678 |
| C | -0.67949976 | -4.24589637 | 3.06682634 |
| C | 2.42386108 | -2.43730389 | 3.12319923 |
| C | 2.89529289 | -1.12593733 | 3.16284588 |
| C | 2.89502759 | 1.12622942 | 3.16291912 |
| C | 4.25975324 | 0.68682053 | 3.10006405 |
| C | 4.25989686 | -0.68625336 | 3.09996465 |
| C | 2.42348856 | 2.43757151 | 3.12369366 |
| C | 1.08391326 | 2.84684258 | 3.13683352 |
| C | 0.67964507 | 4.24655336 | 3.06772913 |
| C | -0.67953947 | 4.24670293 | 3.06778275 |
| C | -1.08408995 | 2.84707882 | 3.13702574 |
| H | -5.11059979 | -1.35610637 | 3.03249610 |
| H | -5.11090407 | 1.35668751 | 3.03257384 |
| H | -3.18177209 | -3.21823863 | 3.06722577 |
| H | 1.35970804 | -5.09152882 | 3.00925464 |
| H | -1.35968196 | -5.09126645 | 3.00951357 |
| H | 3.18225201 | -3.21864311 | 3.06675202 |
| H | 5.11098105 | -1.35600804 | 3.03196633 |
| H | 5.11068060 | 1.35675893 | 3.03207784 |
| H | 3.18183601 | 3.21892982 | 3.06714186 |
| H | 1.35982713 | 5.09189641 | 3.01037225 |
| H | -1.35956441 | 5.09220806 | 3.01061368 |
| H | -3.18203804 | 3.21938855 | 3.06766104 |
| H | -1.09431758 | 0.00085733 | 3.25882458 |
| H | 1.09446134 | -0.00016673 | 3.25866765 |

**Table S10. DFT calculated geometry of $F^{-1}$ on 4 ML NaCl. All values reported are in Å.**

| | | | |
|---|---:|---:|---:|
| N | -2.17572336 | 0.02040935 | 3.13618948 |
| N | -0.00260467 | -1.94492787 | 2.99568117 |
| N | 2.17059372 | 0.02035437 | 3.13555301 |
| N | -0.00214960 | 1.98521078 | 3.01107592 |
| C | -4.30454395 | 0.71816046 | 2.98397211 |
| C | -4.30436577 | -0.67720400 | 2.98291941 |
| C | -2.95720646 | -1.11043482 | 3.06229595 |
| C | -2.95750401 | 1.15155425 | 3.06522173 |
| C | -2.43946756 | 2.46150935 | 3.01385654 |
| C | -2.44042384 | -2.42105426 | 3.00918677 |
| C | -1.09653598 | -2.78688483 | 2.99660671 |
| C | 1.09016135 | -2.78688259 | 2.99919043 |
| C | 0.67515218 | -4.18516935 | 2.98471929 |
| C | -0.68151186 | -4.18534324 | 2.98276389 |
| C | 2.43428565 | -2.42137776 | 3.01222164 |
| C | 2.95156671 | -1.11059815 | 3.06393892 |
| C | 2.95234531 | 1.15025444 | 3.06532545 |
| C | 4.29923411 | 0.71706096 | 2.98597882 |
| C | 4.29834580 | -0.67849140 | 2.98527773 |
| C | 2.43549803 | 2.46073365 | 3.01471604 |
| C | 1.09132170 | 2.82585217 | 3.00488324 |
| C | 0.67645545 | 4.22493781 | 2.98010391 |
| C | -0.67998090 | 4.22488827 | 2.98014602 |
| C | -1.09552545 | 2.82621486 | 3.00326456 |
| H | -5.15694093 | -1.34360736 | 2.89891045 |
| H | -5.15701666 | 1.38469754 | 2.90155078 |
| H | -3.17311118 | -3.22799626 | 2.98562661 |
| H | 1.35631791 | -5.03258771 | 2.97341736 |
| H | -1.36299032 | -5.03234136 | 2.97097436 |
| H | 3.16691549 | -3.22874125 | 2.99014765 |
| H | 5.15071156 | -1.34470901 | 2.90252331 |
| H | 5.15179337 | 1.38253571 | 2.90361700 |
| H | 3.16728955 | 3.26827472 | 2.98779238 |
| H | 1.35719680 | 5.07152310 | 2.96227582 |
| H | -1.36061124 | 5.07197236 | 2.96134695 |
| H | -3.17093925 | 3.26964162 | 2.98639839 |
| H | -1.17041471 | 0.02262208 | 3.30280364 |
| H | 1.16519696 | 0.02261673 | 3.30191198 |

**Table S11. DFT calculated geometry of $F^{-2}$ on 4 ML NaCl. All values reported are in Å.**

| | | | |
|---|---|---|---|
| N | -2.12441922 | 0.00019333 | 3.09577593 |
| N | 0.00006466 | -2.03268562 | 3.02678108 |
| N | 2.12464139 | 0.00019087 | 3.09555667 |
| N | 0.00008713 | 2.03295208 | 3.02678403 |
| C | -4.28865606 | 0.68417476 | 2.96331999 |
| C | -4.28865763 | -0.68401089 | 2.96344891 |
| C | -2.92246299 | -1.14597364 | 3.05081557 |
| C | -2.92267244 | 1.14635075 | 3.05062212 |
| C | -2.46118885 | 2.43911921 | 3.02255997 |
| C | -2.46106322 | -2.43880124 | 3.02267632 |
| C | -1.09441429 | -2.87334067 | 3.03031437 |
| C | 1.09465638 | -2.87330977 | 3.03017446 |
| C | 0.69824738 | -4.23915626 | 3.02494311 |
| C | -0.69805866 | -4.23912131 | 3.02497133 |
| C | 2.46134298 | -2.43877775 | 3.02225836 |
| C | 2.92276726 | -1.14599220 | 3.05030113 |
| C | 2.92279293 | 1.14627271 | 3.05041687 |
| C | 4.28889397 | 0.68424992 | 2.96316790 |
| C | 4.28885122 | -0.68396593 | 2.96311468 |
| C | 2.46134754 | 2.43913086 | 3.02240081 |
| C | 1.09466125 | 2.87354418 | 3.03027243 |
| C | 0.69827037 | 4.23937793 | 3.02511502 |
| C | -0.69802616 | 4.23938230 | 3.02526401 |
| C | -1.09453428 | 2.87350482 | 3.03042394 |
| H | -5.14231561 | -1.34970632 | 2.87132309 |
| H | -5.14241530 | 1.34976020 | 2.87107203 |
| H | -3.22052442 | -3.22271433 | 3.00211264 |
| H | 1.37279419 | -5.09282905 | 3.04047524 |
| H | -1.37258020 | -5.09285599 | 3.04075160 |
| H | 3.22082839 | -3.22261110 | 3.00160417 |
| H | 5.14256738 | -1.34955227 | 2.87097153 |
| H | 5.14252542 | 1.34993438 | 2.87096637 |
| H | 3.22085475 | 3.22294463 | 3.00175684 |
| H | 1.37280848 | 5.09303428 | 3.04083921 |
| H | -1.37243419 | 5.09311623 | 3.04090699 |
| H | -3.22062777 | 3.22299347 | 3.00175163 |
| H | -1.19971267 | 0.00035094 | 3.51620806 |
| H | 1.19998493 | 0.00018394 | 3.51623879 |